\documentclass[11pt]{article} 
\usepackage[left=2.5cm, right=2.5cm, top=2.5cm, bottom=2.5cm]{geometry}

\usepackage{amssymb}
\usepackage{amsmath}

\usepackage[colorinlistoftodos]{todonotes}

\usepackage{listings} 
\usepackage{showexpl} 
\usepackage{color} 
\usepackage{amsthm} 
\usepackage{alltt}
\usepackage[export]{adjustbox} 

\usepackage[ruled,vlined]{algorithm2e}
\usepackage{algpseudocode}
\usepackage{amssymb}
\usepackage{amsmath}
\usepackage{amsfonts}
\usepackage{booktabs}
\usepackage{enumerate}
\usepackage{fancyhdr}
\usepackage{float}
\usepackage{graphicx}
\usepackage{lastpage}
\usepackage{mathrsfs}
\usepackage{authblk}
\usepackage{setspace}

\usepackage[colorlinks=true, citecolor=blue]{hyperref}
\hypersetup{
    colorlinks=true,
    linkcolor=blue,
    filecolor=magenta,      
    urlcolor=cyan,
}

\usepackage{appendix}
\usepackage{tikz}
\usetikzlibrary{decorations.text}
\usetikzlibrary{shapes,arrows}
\usepackage{pgfplots}
\usepackage{pgfplotstable}
\pgfplotsset{compat=newest}

\usetikzlibrary{calc}
\usetikzlibrary{spy}
\usetikzlibrary{positioning}
\usepackage{pdftexcmds}
\usetikzlibrary{external}
\usetikzlibrary{positioning}
\usetikzlibrary{arrows}
\usetikzlibrary{decorations.markings}
\usetikzlibrary{decorations.text}
\usetikzlibrary{backgrounds}
\usetikzlibrary{spy}
\usetikzlibrary{calc,patterns,decorations.pathmorphing,decorations.markings}
\usetikzlibrary{decorations.pathreplacing}
\usepgfplotslibrary{patchplots}

\usepackage{chngcntr}
\usepackage{etoolbox}
\usepackage{lipsum}

\usepackage{bm}

\makeatletter
\providecommand*{\input@path}{}
\g@addto@macro\input@path{{./}}
\makeatother

\makeatletter

\def\pgfplotstableread@openfile{%
    \def\pgfplotstable@loc@TMPa{\pgfutil@in@{ }}%
    \expandafter\pgfplotstable@loc@TMPa\expandafter{\pgfplotstableread@filename}%
    \ifpgfutil@in@
        \t@pgfplots@toka=\expandafter{\pgfplotstableread@filename}%
        \edef\pgfplotstableread@filename{\pgfplots@dquote\the\t@pgfplots@toka\pgfplots@dquote}%
    \fi
    \let\pgfplotstableread@old@crcr=\\%
    \def\\{\string\\}
    \openin\r@pgfplots@reada=\csname pgfk@/pgfplots/table file path\endcsname\pgfplotstableread@filename.tex
    \ifeof\r@pgfplots@reada
        \openin\r@pgfplots@reada=\csname pgfk@/pgfplots/table file path\endcsname\pgfplotstableread@filename\relax
    \else
        \pgfplots@warning{%
            You requested to open table '\pgfplotstableread@filename', but there is also a '\pgfplotstableread@filename.tex'. 
            TeX will automatically append the suffix '.tex', so I will now open '\pgfplotstableread@filename.tex'.
            Please make sure you don't accidentally load TeX files - this may produce unrecoverable errors.}%
        \closein\r@pgfplots@reada
        \openin\r@pgfplots@reada=\pgfplotstableread@filename\relax
    \fi
    \ifeof\r@pgfplots@reada
        \pgfplotsthrow{no such table file}{\pgfplots@loc@TMPa}{\pgfplotstableread@filename}{Could not read table file '\csname pgfk@/pgfplots/table file path\endcsname\pgfplotstableread@filename'. In case you intended to provide inline data: maybe TeX screwed up your end-of-lines? Try `row sep=crcr' and terminate your lines with `\string\\' (refer to the pgfplotstable manual for details)}\pgfeov%
        \global\let\pgfplotstable@colnames@glob=\pgfplots@loc@TMPa
        \def\pgfplotstableread@ready{0}%
    \fi
    \pgfplots@logfileopen{\pgfplotstableread@filename}%
    \let\\=\pgfplotstableread@old@crcr
}

\makeatother

\pgfplotscreateplotcyclelist{blue to red}{%
  color=blue\\%
  color=red!10!blue\\%
  color=red!20!blue\\%
  color=red!30!blue\\%
  color=red!40!blue\\%
  color=red!50!blue\\%
  color=red!60!blue\\%
  color=red!70!blue\\%
  color=red!80!blue\\%
  color=red!90!blue\\%
  color=red\\%
}
\pgfplotscreateplotcyclelist{red to blue}{%
  color=red\\%
  color=red!90!blue\\%
  color=red!80!blue\\%
  color=red!70!blue\\%
  color=red!60!blue\\%
  color=red!50!blue\\%
  color=red!40!blue\\%
  color=red!30!blue\\%
  color=red!20!blue\\%
  color=red!10!blue\\%
  color=blue\\%
}

\pgfplotscreateplotcyclelist{red to blue fast}{%
  color=red\\%
  color=red!80!blue\\%
  color=red!60!blue\\%
  color=red!40!blue\\%
  color=red!20!blue\\%
  color=blue\\%
}

\pgfplotscreateplotcyclelist{2red to blue}{%
  color=red\\%
  color=red\\%
  color=red!90!blue\\%
  color=red!90!blue\\%
  color=red!80!blue\\%
  color=red!80!blue\\%
  color=red!70!blue\\%
  color=red!70!blue\\%
  color=red!60!blue\\%
  color=red!60!blue\\%
  color=red!50!blue\\%
  color=red!50!blue\\%
  color=red!40!blue\\%
  color=red!40!blue\\%
  color=red!30!blue\\%
  color=red!30!blue\\%
  color=red!20!blue\\%
  color=red!20!blue\\%
  color=red!20!blue\\%
  color=red!10!blue\\%
  color=blue\\%
  color=blue\\%
}

\pgfplotsset{discard if/.style 2 args={x filter/.code={\ifnum\thisrow{#1}=#2\else\fi}}}

\usepackage{bm}

\newcommand{\vect}[1]{\mathbf{#1}}

\newcommand{\lambdab}{{\boldsymbol \lambda}}

\newcommand{\cs}{\boldsymbol \sigma}

\newcommand{\thetab}{\boldsymbol \theta}

\newcommand{\Fb}{{\boldsymbol F}}
\newcommand{\Hb}{{\boldsymbol H}}
\newcommand{\Ib}{{\boldsymbol I}}

\newcommand{\Pb}{{\boldsymbol P}}

\newcommand{\fb}{{\boldsymbol f}}

\newcommand{\hb}{{\boldsymbol h}}

\newcommand{\nb}{{\boldsymbol n}}

\newcommand{\rb}{{\boldsymbol r}}

\newcommand{\tb}{{\boldsymbol t}}
\newcommand{\ub}{{\boldsymbol u}}
\newcommand{\vb}{{\boldsymbol v}}

\newcommand{\yb}{{\boldsymbol y}}

\newcommand{\x}{{\boldsymbol x}}
\newcommand{\y}{{\boldsymbol y}}

\def\gL{{\mathcal{L}}}

\def\sR{{\mathbb{R}}}

\usepackage{amsmath}

\usepackage{nicefrac}
\usepackage{multirow}
\usepackage{hhline}
\usepackage{lineno}
\usepackage{import}
\usepackage{epstopdf}
\usepackage{subcaption}
\usepackage{mathtools}
\usepackage{transparent}

\usepackage{listings}
\usepackage{cancel}
\usepackage{empheq}
\usepackage{parskip}

\usepackage{abstract}

\definecolor{tbf}{RGB}{255,0,0} 
\definecolor{txue}{RGB}{0,0,255}

\SetKwInput{KwInput}{Input}
\SetKwInput{KwOutput}{Output}

\usepackage{indentfirst}
\title{\Large Implicit differentiation with second-order derivatives and benchmarks in finite-element-based differentiable physics}

\begin{document}

\author[1]{\normalsize Tianju Xue
\footnote{\textit{cetxue@ust.hk} (corresponding author)}}
\affil[1]{\footnotesize Department of Civil and Environmental Engineering, The Hong Kong University of Science and Technology, Hong Kong, China}

\date{}
\maketitle
\vspace{-20pt}

\begin{abstract}

Differentiable programming is revolutionizing computational science by enabling automatic differentiation (AD) of numerical simulations. 
While first-order gradients are well-established, second-order derivatives (Hessians) for implicit functions in finite-element-based differentiable physics remain underexplored. 
This work bridges this gap by deriving and implementing a framework for implicit Hessian computation in PDE-constrained optimization problems. 
We leverage primitive AD tools (Jacobian-vector product/vector-Jacobian product) to build an algorithm for Hessian-vector products and validate the accuracy against finite difference approximations.
Four benchmarks spanning linear/nonlinear, 2D/3D, and single/coupled-variable problems demonstrate the utility of second-order information. 
Results show that the Newton-CG method with exact Hessians accelerates convergence for nonlinear inverse problems (e.g., traction force identification, shape optimization), while the L-BFGS-B method suffices for linear cases. 
Our work provides a robust foundation for integrating second-order implicit differentiation into differentiable physics engines, enabling faster and more reliable optimization.

\end{abstract}

\begin{keywords}
differentiable programming; PDE-constrained optimization; implicit Hessian
\end{keywords}

\section{Introduction}
\label{Sec:Introduction}
 
The advent of differentiable programming—a paradigm shift enabling automatic differentiation (AD) of entire numerical simulations—is revolutionizing computational science. 
Unlike traditional symbolic or numerical differentiation, differentiable programming embeds AD directly into the computational graph of numerical solvers, enabling exact gradient computation without manual derivation~\cite{blondel2024elements}.

\begin{figure}[H] 
\centering
\includegraphics[scale=0.55]{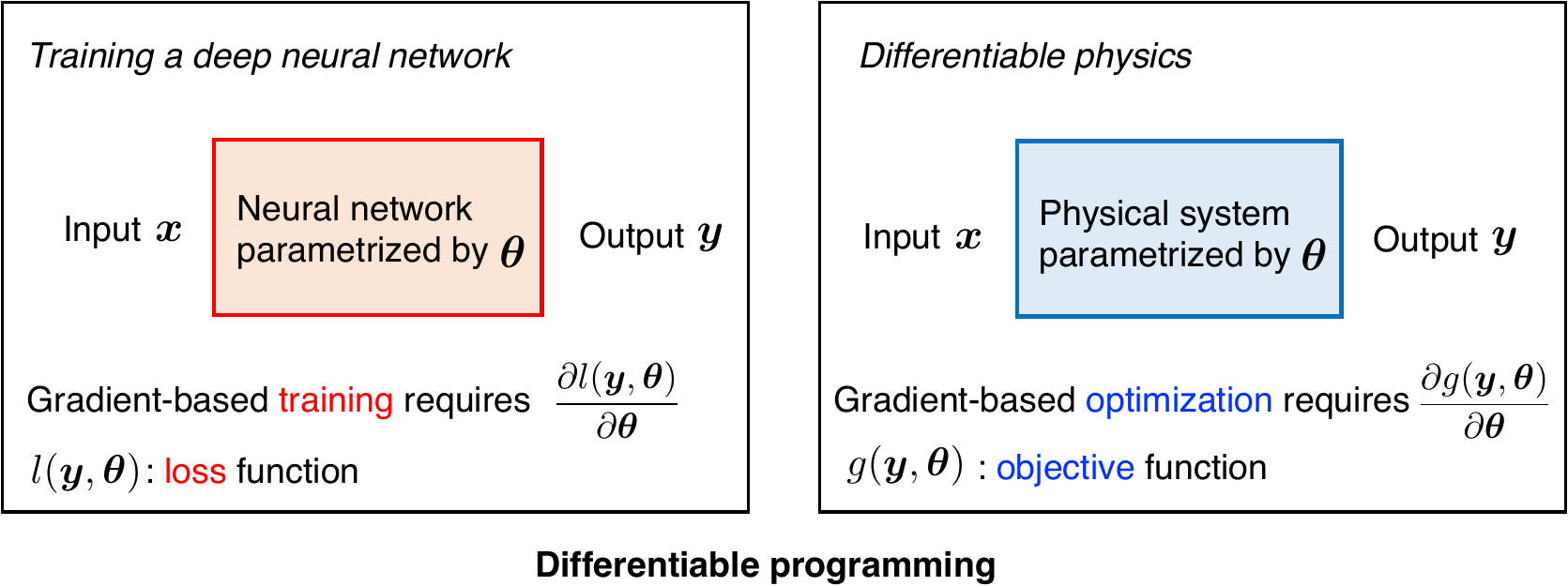}
\caption{Differentiable programming breaks the boundary between deep learning and differentiable physics.}
\label{Fig:differentiable_programming}
\end{figure}

The canonical use of differentiable programming is in the field of deep learning, where the better known ``backpropagation'', i.e., reverse mode of AD, has been the engine of training deep neural networks~\cite{lecun2015deep}.
In the field of computational science, differentiable programming is rapidly rising, since one can equip classic physics-based numerical solvers with differentiable programming and obtain a differentiable physics engine easy for inverse optimization.
As shown in Fig.~\ref{Fig:differentiable_programming}, the side-by-side comparison between training a deep neural network and differentiable physics reveals much similarity between them: both workflows have an input $\x$ and an output $\y$; the forward functions (either a neural network or a numerical solver) are parametrized by $\thetab$; there is a loss function $l(\y,\thetab)$ (or an objective function $g(\y,\thetab)$); the goal is to find desired $\thetab$ by gradient-based training (or optimization) algorithms that require $\frac{\partial l(\y,\thetab)}{\partial \thetab}$ (or $\frac{\partial g(\y,\thetab)}{\partial \thetab}$).
As the heart of both workflows, it is differentiable programming that enables successful automatic computation of the gradients, even for much complicated nonlinear systems.

Frameworks like \texttt{JAX}~\cite{bradbury2018jax} or \texttt{PyTorch}~\cite{paszke2019pytorch} have democratized differentiable programming, catalyzing various applications in differentiable physics across many fields such as fluid mechanics~\cite{bezgin2023jax,kochkov2021machine}, molecular dynamics~\cite{schoenholz2020jax}, solid mechanics~\cite{pundir2025simplifying,du2024neural}, heat transfer~\cite{shang2025jax}, etc.
However, a key technical difference between finding the gradients $\frac{\partial l(\y,\thetab)}{\partial \thetab}$ in deep learning and $\frac{\partial g(\y,\thetab)}{\partial \thetab}$ in differentiable physics is that neural network functions are often explicit while numerical solvers for physical systems are often implicit.
The direct consequence is that the output $\y$ as a function of $\thetab$ and $\x$ is often explicit in deep learning so that AD can be directly applied, while this relation is often implicit in differentiable physics so that naive AD cannot be applied due to internal iterations for solving nonlinear equations.
Most of the previous works on differentiable physics have adopted explicit numerical solvers to bypass this issue. 
The proper way to handle gradients related to implicit functions is to integrate the adjoint method~\cite{cao2003adjoint} into customized AD rules, as extensively discussed in~\cite{blondel2022efficient}.

Based on adjoint-enabled customized AD rules, we have proposed a differentiable finite element method (FEM) in our previous work~\texttt{JAX-FEM}~\cite{xue2023jax}, a differentiable physics engine that can handle implicit functions.
We select finite element methods~\cite{hughes2003finite} as the tool to discretize physical systems since it is one of the most powerful approaches for numerical solutions to partial differential equations (PDEs).
The methodology in principle applies to other numerical methods, such as finite volume methods~\cite{ferziger2019computational} or mesh-free methods~\cite{liu2009meshfree}.
Within the scope of differentiable physics, \texttt{JAX-FEM} effectively handles inverse problems, mathematically formulated as PDE-constrained optimization problems~\cite{rees2010optimal}.

Despite recent progress in first-order differentiable physics enabled by adjoint-based implicit differentiable programming, second-order implicit differentiation, i.e., Hessian matrices, remains underexplored. 
Hessians encode curvature information about the landscape of the objective function, allowing Newton-type algorithms (e.g., Newton-CG~\cite{fletcher1964function,nash1984newton}) to achieve quadratic convergence rates near minima~\cite{biegler1995reduced,nocedal1999numerical}.
The potential advantage of accelerated optimization convergence using the Hessian information is appealing and desired in the differentiable programming community (see issue $\#$474 in the discussion forum of the implicit differentiation package \texttt{JAXopt}~\cite{blondel2022efficient}), yet this Hessian information for implicit differentiation is currently not available.
In general, there are two strategies to solve PDE-constrained optimization problems: optimize-then-discretize~\cite{cipolla2018fractional} and discretize-then-optimize~\cite{betts2005discretize}.
Existing efforts to determine Hessian information mostly follow the optimize-then-discretize strategy~\cite{bui2013computational}, such as the \texttt{hIPPYlib}~\cite{villa2021hippylib} package for Bayesian inverse problems.
However, the workflow of differentiable physics tends to prefer the discretize-then-optimize strategy, since it offers higher flexibility and integrates deeper with AD, showcasing a higher level of automation.

This work bridges the gap between classical adjoint-based Hessian methods and modern differentiable programming by developing a unified and practical framework for second-order implicit differentiation in FEM-based differentiable physics. 
The key contributions are:
\begin{enumerate}
    \item \textit{Derivation}: We follow the discretize-then-optimize strategy and derive the Hessian matrix under the implicit differentiable programming paradigm. 
    Our approach can be considered as a second-order generalization of the first-order framework of Blondel et~\cite{blondel2022efficient}.
    \item  \textit{Implementation}: We implement the Hessian-vector products using \texttt{JAX}'s primitive functions Jacobian-vector product (JVP) and vector-Jacobian product (VJP), and validate the accuracy and compare the efficiency with a finite-difference-based approximate approach.
    \item  \textit{Benchmarking}: We design four benchmark inverse problems (linear/nonlinear, 2D/3D, single variable/coupled variables) that cover several typical tasks: parameter identification, deformation control, and shape optimization.
    Based on the results, we obtain insights on when second-order information is useful for better optimization convergence.
\end{enumerate}

The paper is organized as follows.
Section~\ref{Sec:implicit_diff} defines the concept of implicit differentiation, states a model problem and provides a derivation for the Hessian matrix.
Section~\ref{Sec:implementation} shows the details of implementation of implicit Hessian-vector products.
Section~\ref{Sec:numerical} presents four representative numerical benchmarks to discuss the effectiveness of using Hessian information to help optimization convergence.
We then conclude in Section~\ref{Sec:conclusions}.

Our code is available at \href{https://github.com/tianjuxue/hessian}{https://github.com/tianjuxue/hessian}.

\section{Implicit differentiation}
\label{Sec:implicit_diff}

In this section, we first introduce the concept of implicit differentiation in the differentiable programming context. 
A nonlinear Poisson's problem is considered as a model problem for PDE-constrained optimization, where the discretization is performed using the finite element method. 
With the help of the Lagrange function, we derive the expressions for the first-order derivative (gradient) and the second-order derivative (Hessian) of the objective function with respect to optimization parameters.

\subsection{Problem statement}

\subsubsection{General concept of ``explicit differentiation'' versus ``implicit differentiation''}

Consider a (discretized) physical system governed by parameters $\thetab \in \mathbb{R}^M$, as shown in the right panel of Fig.~\ref{Fig:differentiable_programming}.
The system takes an input $\x \in \mathbb{R}^L$ and generates an output $\y \in \mathbb{R}^N$.
The system can be optimized by adjusting the parameter vector $\thetab$ to minimize a certain objective function $g(\y, \thetab)$ where $g(\cdot, \cdot): \sR^{N}\times \sR^{M} \rightarrow \sR$.
Since $M$ is typically large, such high-dimensional optimization problems are usually solved relying on the gradients, i.e., $\frac{\textrm{d}g}{\textrm{d}\thetab}$.

In the explicit case, the output $\y$ is explicitly evaluated given $\x$ and $\thetab$, i.e., $\y=\hb(\x, \thetab)$ where $\hb(\cdot, \cdot): \sR^{L}\times \sR^{M} \rightarrow \sR^N$.
Therefore, the gradient can be evaluated also explicitly
\begin{align}
    \frac{\textrm{d}g}{\textrm{d}\thetab} =  \frac{\partial g}{\partial \y} \frac{\partial \y}{\partial \thetab} + \frac{\partial g}{\partial \thetab} =  \frac{\partial g}{\partial \y} \frac{\partial \hb}{\partial \thetab} + \frac{\partial g}{\partial \thetab}.
\end{align}
Applying automatic differentiation techniques to find the above gradient is straightforward, the procedure known as \textit{explicit differentiation}.

In the implicit case, the output $\y$ is implicitly related to $\x$ and $\thetab$ through a (nonlinear) residual function $\rb(\x, \y, \thetab)=\textbf{0}$ where $\rb(\cdot,\cdot,\cdot):\sR^L\times\sR^N\times\sR^M \rightarrow \sR^N$.
Therefore, $\y=\y(\x, \thetab)$ can be only be evaluated by solving a set of nonlinear equations, and the gradient
\begin{align}
    \frac{\textrm{d}g}{\textrm{d}\thetab} =  \frac{\partial g}{\partial \y} \frac{\partial \y}{\partial \thetab} + \frac{\partial g}{\partial \thetab} = \cdots (\textrm{extra procedures, must deal with $\frac{\partial \y}{\partial \thetab}$})
\end{align}
should be evaluated with extra steps with proper substitution of $\frac{\partial \y}{\partial \thetab}$, the procedure known as \textit{implicit differentiation}.

Problems requiring implicit differentiation are ubiquitous. 
Blondel et al.~\cite{blondel2022efficient} discussed the situations involving implicit differentiation and the implementation using the modern automatic differentiation architecture \texttt{JAX}.
In our previous work \texttt{JAX-FEM}~\cite{xue2023jax}, the implicit differentiation methods have been developed to solve inverse problems discretized by finite element methods.
However, limited research has focused on second-order information, i.e., Hessian $\frac{\textrm{d}^2 g}{\textrm{d}\thetab^2}$, which is the central topic of this work.
In what follows, we will present a model problem in the finite element context, and endow $\x$, $\y$, $\thetab$, $\rb(\cdot,\cdot,\cdot)$ and $g(\cdot,\cdot)$ with specific physical meanings.

\subsubsection{A model problem in the finite element context}
\label{Sec:model_problem}

Although implicit differentiation occurs in many situations, we will focus on inverse problems discretized by finite element methods for this work.
To illustrate the problem setup, let us consider a typical inverse problem of parameter identification under the constraint of a nonlinear Poisson's equation.
We will first describe the forward prediction problem where the unknown solution vector is to be sought given the parameter vector, and then the inverse identification problem where the parameter vector is to be sought given the observed solution vector.

For the forward prediction, the strong form of the nonlinear Poisson's problem states that find $u: \Omega \rightarrow \mathbb{R}$ such that
\begin{align} \label{Eq:strong}
    -\nabla \cdot \big( \textrm{exp}(\theta \, u) \, \nabla u \big) = b & \quad \textrm{in}  \, \, \Omega, \nonumber \\
    u = u_D &  \quad\textrm{on} \, \, \Gamma_D,  \nonumber \\
   \textrm{exp}(\theta \, u) \, \nabla u \cdot \nb = t  & \quad \textrm{on} \, \, \Gamma_N.
\end{align}
where $\Omega\subset\mathbb{R}^{\textrm{dim}}$ is the problem domain, $\theta:\Omega\rightarrow\sR$ is the parameter field, $b$ is the source term, $u_D$ is the Dirichlet boundary condition defined on ${\Gamma_D\subset\partial\Omega}$, $\nb$ is the outward normal, and $t$ prescribes Neumann boundary condition on ${\Gamma_N\subset\partial\Omega}$ (${\Gamma_D\cup\Gamma_N=\partial\Omega}$ and ${\Gamma_D\cap\Gamma_N=\emptyset}$).

The weak form of Eq.~(\ref{Eq:strong}) states the following: Find $u\in\mathcal{U}$ such that $\forall v \in \mathcal{V}$
\begin{align} \label{Eq:weak}
  r(u; v) = \int_{\Omega}  \textrm{exp}(\theta \, u) \, \nabla u  \cdot  \nabla v \textrm{ d} \Omega - \int_{\Gamma_N} t \,  v \textrm{ d} \Gamma - \int_{\Omega} b \,  v \textrm{ d} \Omega = 0,
\end{align}
where the trial and test function spaces are 
\begin{align}
    \mathcal{U} &= \big\lbrace u \in H^1(\Omega) \,\, \big| \,\, u = u_D \textrm{ on } \Gamma_D \big\rbrace,  \nonumber \\
    \mathcal{V} &= \big\lbrace v \in H^1(\Omega) \,\, \big| \,\, v = 0 \textrm{ on } \Gamma_D \big\rbrace.
\end{align}

The finite element approximation requires that
\begin{align} \label{Eq:fem_weak}
  r(u^h; v^h) = \int_{\Omega}  \textrm{exp}(\theta \, u^h) \, \nabla u^h  \cdot  \nabla v^h \textrm{ d} \Omega - \int_{\Gamma_N} t \,  v^h \textrm{ d} \Gamma - \int_{\Omega} b \,  v^h \textrm{ d} \Omega = 0,
\end{align}
where the trial and test function spaces are 
\begin{align}
   \mathcal{U} \supset \mathcal{U}^h &= \big\lbrace u^h \in H^1(\Omega) \,\, \big| \,\, u^h(\x) = \sum_{i=1}^N u_i \, \phi_i(\x), \, u^h = u_D \textrm{ on } \Gamma_D \big\rbrace,  \nonumber \\
   \mathcal{V} \supset \mathcal{V}^h &= \big\lbrace v^h \in H^1(\Omega) \,\, \big| \,\, v^h(\x) = \sum_{i=1}^N v_i \, \phi_i(\x), \, v^h = 0 \textrm{ on } \Gamma_D \big\rbrace,
\end{align}
where the approximate solution $u^h$ is constructed with nodal values $u_i$ and shape functions $\phi_i$.
Let $\y=[u_1, u_2,\dots,u_N]$ be the collection of degrees of freedom, $\x=[b_1,b_2,\dots,b_L]$ be the discretized version of the source term, and $\thetab=[\theta_1,\theta_2,\dots,\theta_M]$ be the parameter vector.
Then, the solution vector $\y$ of this nonlinear Poisson's problem is related to $\x$ and $\thetab$ in an implicit way by solving the residual equation arising from Eq.~(\ref{Eq:fem_weak}):
\begin{align}
    \rb(\x,\y,\thetab)=\textbf{0}.
\end{align}

The inverse parameter identification is formulated as a PDE-constrained optimization problem:
\begin{align} \label{Eq:PDE-CO}
    \nonumber &\min_{u,\theta}  \frac{1}{2}\int_{\Omega}  (u - u_{\textrm{obs}})^2 \textrm{ d} \Omega
        + \frac{\alpha}{2} \int_{\Omega} \theta^2 \textrm{ d} \Omega \\
    & \qquad \qquad \textrm{s.t.} \quad r(u;v)=0, 
\end{align}
where the objective functional $g(u,\theta)$ contains a data mismatch term between the predicted solution $u$ and the observation $u_{\textrm{obs}}$, and a regularization term controlled by $\alpha$.
The optimization problem is constrained by the weak form residual from Eq.~(\ref{Eq:weak}).

The discretized version of problem (\ref{Eq:PDE-CO}) is 
\begin{align} \label{Eq:PDE-CO_discretized}
    &\nonumber \min_{\y \in\sR^{N}, \thetab\in\sR^{M}} g(\y, \thetab) \\
    & \textrm{s.t.} \quad \rb(\x, \y, \thetab)=\boldsymbol{0}.
\end{align}

To solve this high-dimensional optimization problem,  derivative information is crucial. 
In the following subsections, we will show the derivation of the first-order derivative $\frac{\textrm{d}g}{\textrm{d}\thetab}$ and the second-order derivative $\frac{\textrm{d}^2 g}{\textrm{d}\thetab^2}$ under the implicit differentiation context.

\subsection{First-order derivative: The gradient}

The goal here is to find the first-order derivative of the objective function $g$ to $\thetab$, i.e., $\frac{\textrm{d}g}{\textrm{d}\thetab}\in \mathbb{R}^M$. 
We define the Lagrangian function
\begin{align} \label{Eq:lagrangian}
\mathcal{L}(\y, \lambdab, \vect{\thetab}) = g(\y, \vect{\thetab}) + \lambdab^T \rb(\y, \vect{\thetab}),
\end{align}
where $\lambdab \in \mathbb{R}^N$ is the Lagrange multiplier vector (also the adjoint vector), and we have omitted the dependence of the residual $\rb$ on the input vector $\x$ in the notation since we are not concerned with the derivative of $\rb$ with respect to $\x$. 
The  stationary points $(\y, \lambdab)$ satisfy the first-order conditions:
\begin{align} 
    \label{Eq:FOC_adjoint}  
    \frac{\partial\mathcal{L}}{\partial \boldsymbol{y}} &= \frac{\partial g}{\partial \boldsymbol{y}} + \boldsymbol{\lambda}^T \frac{\partial \boldsymbol{r}}{\partial \boldsymbol{y}} = \boldsymbol{0} \\  
    \label{Eq:FOC_fwd} 
    \frac{\partial\mathcal{L}}{\partial \boldsymbol{\lambda}} &= \boldsymbol{r}(\boldsymbol{y}, \boldsymbol{\theta}) = \boldsymbol{0},
\end{align}
where the adjoint equation and the residual equation have been recovered.
It is straightforward to show that the total gradient $\frac{\textrm{d} g}{\textrm{d} \thetab}$ is given by the partial derivative of $\gL$ to $\thetab$:
\begin{align} \label{Eq:gradient}
    \frac{\textrm{d} g}{\textrm{d} \thetab} = \frac{\partial \gL}{\partial \thetab} = \frac{\partial g}{\partial \thetab} + \lambdab^T \frac{\partial \rb}{\partial \thetab}.
\end{align}

\subsection{Second-order derivative: The Hessian}

The goal here is to find the second-order derivative of the objective function $g$ to $\thetab$, i.e., $\frac{\textrm{d}^2 g}{\textrm{d}\thetab^2}$.
We differentiate the first-order conditions (\ref{Eq:FOC_adjoint}) and (\ref{Eq:FOC_fwd}) with respect to $\thetab$ and apply the chain rule to obtain:
\begin{align} \label{Eq:incremental}
\begin{bmatrix}
   \frac{\partial}{\partial \y}\frac{\partial \gL}{\partial \y}  & \frac{\partial}{\partial \lambdab}\frac{\partial \gL}{\partial \y} \\
    \frac{\partial}{\partial \y}\frac{\partial \gL}{\partial \lambdab}  & \frac{\partial}{\partial \lambdab}\frac{\partial \gL}{\partial \lambdab}
\end{bmatrix}
\begin{bmatrix}
    \frac{\partial \y}{\partial \vect{\thetab}} \\
    \frac{\partial \lambdab}{\partial \vect{\thetab}}
\end{bmatrix}
= -
\begin{bmatrix}
   \frac{\partial}{\partial \thetab} \frac{\partial \gL}{\partial \y}\\ 
   \frac{\partial}{\partial \thetab} \frac{\partial \gL}{\partial \lambdab} 
\end{bmatrix}. 
\end{align}

The Hessian matrix $\Hb\in\sR^{M\times M}$ is given by differentiating the gradient in Eq.~(\ref{Eq:gradient}) with respect to $\thetab$ and apply the chain rule:
\begin{align} \label{Eq:hessian}
    \nonumber
    \Hb=\frac{\textrm{d} }{\textrm{d} \thetab} \frac{\textrm{d} g}{\textrm{d}\thetab} &= \frac{\partial }{\partial\thetab} \frac{\partial \gL}{\partial\thetab} + 
    \begin{bmatrix} 
    \frac{\partial}{\partial \y} \frac{\partial \gL}{\partial \thetab} & \frac{\partial}{\partial \lambdab} \frac{\partial \gL}{\partial \thetab}
    \end{bmatrix} 
    \begin{bmatrix}
    \frac{\partial \y}{\partial \vect{\thetab}} \\
    \frac{\partial \lambdab}{\partial \vect{\thetab}}
    \end{bmatrix} \\
    &=
    \frac{\partial }{\partial\thetab} \frac{\partial \gL}{\partial\thetab} -
    \begin{bmatrix} 
    \frac{\partial}{\partial \y} \frac{\partial \gL}{\partial \thetab} & \frac{\partial}{\partial \lambdab} \frac{\partial \gL}{\partial \thetab}
    \end{bmatrix}
    \begin{bmatrix}
    \frac{\partial}{\partial \y}\frac{\partial \gL}{\partial \y}  & \frac{\partial}{\partial \lambdab}\frac{\partial \gL}{\partial \y} \\
    \frac{\partial}{\partial \y}\frac{\partial \gL}{\partial \lambdab}  & \frac{\partial}{\partial \lambdab}\frac{\partial \gL}{\partial \lambdab}
    \end{bmatrix}^{-1}
    \begin{bmatrix}
   \frac{\partial}{\partial \thetab} \frac{\partial \gL}{\partial \y}\\ 
   \frac{\partial}{\partial \thetab} \frac{\partial \gL}{\partial \lambdab} 
    \end{bmatrix},
\end{align}
where the symmetric property of the Hessian matrix becomes obvious.

The terms in Eq.~(\ref{Eq:hessian}) can be all be found by using the definition of Lagrangian in Eq.~(\ref{Eq:lagrangian}).
The partial second-order term is given by
\begin{align}
    \frac{\partial }{\partial\thetab} \frac{\partial \gL}{\partial\thetab} = \frac{\partial}{\partial \thetab}\frac{\partial g}{\partial \thetab} + \frac{\partial}{\partial \thetab} \Big( \lambdab^T\frac{\partial \rb}{\partial \thetab} \Big)\in \mathbb{R}^{M \times M};
\end{align}
the block matrix (known as the bordered Hessian matrix) is given by
\begin{align}
    \begin{bmatrix}
    \frac{\partial}{\partial \y}\frac{\partial \gL}{\partial \y}  & \frac{\partial}{\partial \lambdab}\frac{\partial \gL}{\partial \y} \\
    \frac{\partial}{\partial \y}\frac{\partial \gL}{\partial \lambdab}  & \frac{\partial}{\partial \lambdab}\frac{\partial \gL}{\partial \lambdab}
    \end{bmatrix}
    =
    \begin{bmatrix}
    \frac{\partial }{\partial \y}\frac{\partial g}{\partial \y} + \frac{\partial}{\partial \y} \Big(\lambdab^T\frac{\partial \rb}{\partial \y} \Big) & \frac{\partial \rb}{\partial \y}^T \\
    \frac{\partial \rb}{\partial \y} & \vect{0}
    \end{bmatrix} \in \mathbb{R}^{(N+N) \times (N+N)},
\end{align}
and the block vector is given by
\begin{align}
    \begin{bmatrix} 
    \frac{\partial}{\partial \y} \frac{\partial \gL}{\partial \thetab} & \frac{\partial}{\partial \lambdab} \frac{\partial \gL}{\partial \thetab}
    \end{bmatrix}^T
    =
    \begin{bmatrix}
   \frac{\partial}{\partial \thetab} \frac{\partial \gL}{\partial \y}\\ 
   \frac{\partial}{\partial \thetab} \frac{\partial \gL}{\partial \lambdab} 
    \end{bmatrix}
    = 
    \begin{bmatrix}
    \frac{\partial }{\partial \thetab} \frac{\partial g}{\partial \y} + \frac{\partial }{\partial \thetab}\Big( \lambdab^T\frac{\partial \rb}{\partial \y} \Big) \\
    \frac{\partial \rb}{\partial \thetab} 
    \end{bmatrix} \in \mathbb{R}^{(N+N) \times M}.
\end{align}

The computationally expensive part lies in solving Eq.~(\ref{Eq:incremental}), including the incremental forward problem (solving for $\frac{\partial \y}{\partial \thetab}$) and the incremental adjoint problem (solving for $\frac{\partial \lambdab}{\partial \thetab}$):
\begin{align} 
    \label{Eq:incremental_fwd}  
    \frac{\partial \rb}{\partial \y} \frac{\partial \y}{\partial \thetab} &= -\frac{\partial \rb}{\partial \thetab}\\
    \label{Eq:incremental_adjoint}
    \Big(\frac{\partial \rb}{\partial \y}\Big)^T \frac{\partial \lambdab}{\partial \thetab} &=- \Big(\frac{\partial }{\partial \y}\frac{\partial g}{\partial \y} + \frac{\partial}{\partial \y} \Big(\lambdab^T\frac{\partial \rb}{\partial \y} \Big)\Big) \frac{\partial \y}{\partial \thetab} -   \frac{\partial }{\partial \thetab} \frac{\partial g}{\partial \y} - \frac{\partial }{\partial \thetab}\Big( \lambdab^T\frac{\partial \rb}{\partial \y} \Big),
\end{align}
where in the first line of the incremental forward problem, an $N\times N$ linear system must be solved for $M$ times to obtain $\frac{\partial \y}{\partial \thetab}$, and in the second line of the incremental adjoint problem, another $N\times N$ linear system must be solved for $M$ times to obtain $\frac{\partial \lambdab}{\partial \thetab}$.
Once $\frac{\partial \y}{\partial \thetab}$ and $\frac{\partial \lambdab}{\partial \thetab}$ are available, they can be substituted into Eq.~(\ref{Eq:hessian}) to obtain the Hessian matrix $\Hb$.

\section{Implementation}
\label{Sec:implementation}

In this section, we continue the derivation for the Hessian information by considering Hessian-vector products.
We show the implementation of Hessian-vector products and introduce three ways of AD mode composition.
The accuracy is verified by comparison with the finite difference approach and also using the Taylor remainder test.
The efficiency is discussed for the three ways of implementation.

\subsection{Implicit Hessian-vector product}

Practically, we may not need the full Hessian matrix, but only the Hessian-vector product.
If indeed the full Hessian matrix is needed, it can be built column by column using the Hessian-vector product.
From Eq.~(\ref{Eq:hessian}), we have
\begin{align} \label{Eq:hessp}
    \nonumber
    \Hb\hat{\thetab} 
     &= 
     \frac{\partial }{\partial\thetab} \frac{\partial \gL}{\partial\thetab} \hat{\thetab} + 
    \begin{bmatrix} 
    \frac{\partial}{\partial \y} \frac{\partial \gL}{\partial \thetab} & \frac{\partial}{\partial \lambdab} \frac{\partial \gL}{\partial \thetab}
    \end{bmatrix} 
    \begin{bmatrix}
    \frac{\partial \y}{\partial \vect{\thetab}} \hat{\thetab} \\
    \frac{\partial \lambdab}{\partial \vect{\thetab}} \hat{\thetab}
    \end{bmatrix} \\
    \nonumber
    &=
    \frac{\partial }{\partial\thetab} \frac{\partial \gL}{\partial\thetab} \hat{\thetab} + 
    \begin{bmatrix} 
    \frac{\partial}{\partial \y} \frac{\partial \gL}{\partial \thetab} & \frac{\partial}{\partial \lambdab} \frac{\partial \gL}{\partial \thetab}
    \end{bmatrix} 
    \begin{bmatrix}
    \hat{\y} \\
    \hat{\lambdab}
    \end{bmatrix} \\
    \nonumber
    & = 
    \frac{\partial}{\partial \thetab}\frac{\partial g}{\partial \thetab} \hat{\thetab} + \frac{\partial}{\partial \thetab} \Big( \lambdab^T\frac{\partial \rb}{\partial \thetab} \Big) \hat{\thetab} +     \begin{bmatrix}
    \frac{\partial }{\partial \thetab} \frac{\partial g}{\partial \y} + \frac{\partial }{\partial \thetab}\Big( \lambdab^T\frac{\partial \rb}{\partial \y} \Big) \\
    \frac{\partial \rb}{\partial \thetab} 
    \end{bmatrix}^T   
    \begin{bmatrix}
    \hat{\y} \\
    \hat{\lambdab}
    \end{bmatrix}\\
    & = 
    \frac{\partial}{\partial \thetab}\frac{\partial g}{\partial \thetab} \hat{\thetab} + \frac{\partial}{\partial \thetab} \Big( \lambdab^T\frac{\partial \rb}{\partial \thetab} \Big) \hat{\thetab} + \frac{\partial}{\partial \yb}\frac{\partial g}{\partial \thetab} \hat{\y} + \frac{\partial}{\partial \y}\Big( \lambdab^T\frac{\partial \rb}{\partial \thetab} \Big) \hat{\y} + 
    \hat{\lambdab}^T\frac{\partial \rb}{\partial \thetab}, 
\end{align}
where $\hat{\thetab}\in\sR^M$ is an arbitrarily given incremental parameter vector, $\hat{\y}:= \frac{\partial \y}{\partial \vect{\thetab}} \hat{\thetab}$ is the incremental state vector, and $\hat{\lambdab}:=\frac{\partial \lambdab}{\partial \vect{\thetab}} \hat{\thetab}$ is the incremental adjoint vector.
Accordingly, the incremental forward problem in Eq.~(\ref{Eq:incremental_fwd}) and the incremental adjoint problem in Eq.~(\ref{Eq:incremental_adjoint}) can be modified as
\begin{align} 
    \label{Eq:incremental_fwd_hessp}  
    \frac{\partial \rb}{\partial \y} \hat{\y} &= -\frac{\partial \rb}{\partial \thetab} \hat{\thetab}\\
    \label{Eq:incremental_adjoint_hessp}
    \Big(\frac{\partial \rb}{\partial \y}\Big)^T \hat{\lambdab} &=- \Big(\frac{\partial }{\partial \y}\frac{\partial g}{\partial \y} + \frac{\partial}{\partial \y} \Big(\lambdab^T\frac{\partial \rb}{\partial \y} \Big)\Big) \hat{\y} -    \Big(\frac{\partial }{\partial \thetab} \frac{\partial g}{\partial \y} + \frac{\partial }{\partial \thetab}\Big( \lambdab^T\frac{\partial \rb}{\partial \y} \Big)\Big)\hat{\thetab}.
\end{align}

The complete workflow of computing the Hessian-vector product is shown in Alg.~(\ref{Alg:hessp}).

\begin{algorithm}[H]
\caption{Procedures to compute the implicit Hessian-vector product}\label{Alg:hessp}
\tcp{Given the parameter vector $\thetab$ at which the Hessian-vector product should be evaluated and the incremental parameter vector $\hat{\thetab}$}\
\KwInput{$\thetab$, $\hat{\thetab}$} 
\tcp{Solve the forward problem}
\paragraph{Step 1} Compute the state vector $\y$ by solving Eq.~(\ref{Eq:FOC_fwd}) \\
\tcp{Solve the adjoint problem} 
\paragraph{Step 2} Compute the adjoint vector $\lambdab$ by solving Eq.~(\ref{Eq:FOC_adjoint}) \\
\tcp{Solve the incremental forward problem}
\paragraph{Step 3} Compute the incremental state vector $\hat{\y}$ by solving Eq.~(\ref{Eq:incremental_fwd_hessp}) \\ 
\tcp{Solve the incremental adjoint problem}
\paragraph{Step 4} Compute the incremental adjoint vector $\hat{\lambdab}$ by solving Eq.~(\ref{Eq:incremental_adjoint_hessp}) \\
\tcp{Find the Hessian-vector product}
\paragraph{Step 5} Compute the Hessian-vector product $\Hb\hat{\thetab}$ using Eq.~(\ref{Eq:hessp}) \\
\tcp{Return the Hessian-vector product} 
\KwOutput{$\Hb\hat{\thetab}$ }
\end{algorithm}



\subsection{Building blocks of AD: JVP and VJP functions}

Naive AD algorithms can handle the Hessian-vector product for an explicit function, but fail to handle an implicit function (e.g. $g(\thetab)$ as in our case).
Instead, customized differentiation rules following Alg.~(\ref{Alg:hessp}) must be defined.
To implement such implicit Hessian-vector products, we will rely on two fundamental functions provided by the AD tools: Jacobian-vector product (JVP) and vector-Jacobian product (VJP), which serve as the building blocks~\cite{blondel2024elements}.

Assume $\fb:\sR^M\rightarrow\sR^N$ is a vector-valued function. 
The JVP function: $\sR^M\rightarrow\sR^N$ defines a linear vector-valued function (in $\vb_t$):
\begin{align}\label{Eq:jvp}
    \textrm{JVP}(\x)[\vb_t] = \frac{\partial \fb}{\partial \x}\vb_t,
\end{align}
where $\vb_t\in\sR^M$ is the tangent vector.
The JVP function is the building block for forward-mode AD algorithms.
The VJP function: $\sR^N\rightarrow\sR^M$ defines a linear vector-valued function (in $\vb_c$):
\begin{align}\label{Eq:jvp}
    \textrm{VJP}(\x)[\vb_c] = \boldsymbol{\vb_c}^T\frac{\partial \fb}{\partial \x},
\end{align}
where $\vb_c\in\sR^N$ is the cotangent vector.
The VJP function is the building block for reverse-mode AD algorithms.
In the AD tool~\texttt{JAX}~\cite{bradbury2018jax}, the floating point operations (FLOPs) costs of both the JVP and VJP functions are about three times the cost of just evaluating the function $\fb$.
We can then regard that both the JVP and VJP functions have about the same marginal cost as the function $\fb$.
It must be emphasized that the dimensions $M$ and $N$ here are flattened dimensions. 
In general, the ``vector'' (e.g., $\vb_c$) can be a scalar, a vector, a matrix or any data container that can be flattened. 

The first-order terms appearing in our workflow of implicit Hessian-vector product evaluation (see Alg.~(\ref{Alg:hessp})) can be directly identified as having JVP or VJP structures.
For example, the term 
$\frac{\partial \rb}{\partial \thetab} \hat{\thetab}$ in Eq.~(\ref{Eq:incremental_fwd_hessp}) used in the incremental forward problem is clearly a JVP; the term $\hat{\lambdab}^T\frac{\partial \rb}{\partial \thetab}$ in Eq.~(\ref{Eq:hessp}) used in the final step is clearly a VJP.

The second-order terms appearing in Alg.~(\ref{Alg:hessp}) may be interpreted as a composition of JVP and/or VJP functions.
For example, the term $\Big(\frac{\partial }{\partial \thetab} \frac{\partial g}{\partial \y}\Big)\hat{\thetab}$ in Eq.~(\ref{Eq:incremental_adjoint_hessp}) used in the incremental adjoint problem can be interpreted in three possible ways of composition:
\begin{align*}
\frac{\partial \Big(1 \frac{\partial g}{\partial \y} \Big)}{\partial \thetab} \hat{\thetab} \quad \sim \quad \textrm{JVP over VJP (forward over reverse)} \\
1\frac{\partial \Big( \frac{\partial g}{\partial \thetab} \hat{\thetab}\Big)}{\partial \y} \quad \sim \quad \textrm{VJP over JVP (reverse over forward)} \\
\hat{\thetab}^T\frac{\partial \Big( 1\frac{\partial g}{\partial \thetab} \Big)}{\partial \y}  \quad \sim \quad \textrm{VJP over VJP (reverse over reverse)}
\end{align*}
Here, we have placed ``1'' as the ``vector'' in some of the VJP functions when appropriate.

Similarly, as another example, the term
$\Big(\frac{\partial }{\partial \thetab}\Big( \lambdab^T\frac{\partial \rb}{\partial \y} \Big)\Big)\hat{\thetab}$ in Eq.~(\ref{Eq:incremental_adjoint_hessp}) can also be interpreted in three possible ways of composition:
\begin{align*}
\frac{\partial \Big( \lambdab^T\frac{\partial \rb}{\partial \y} \Big)}{\partial \thetab} \hat{\thetab} \quad \sim \quad \textrm{JVP over VJP (forward over reverse)}\\
\lambdab^T\frac{\partial \Big( \frac{\partial \rb}{\partial \thetab} \hat{\thetab}\Big)}{\partial \y}\quad \sim \quad \textrm{VJP over JVP (reverse over forward)} \\ \hat{\thetab}^T\frac{\partial \Big( \lambdab^T\frac{\partial \rb}{\partial \thetab} \Big)}{\partial \y}   \quad \sim \quad \textrm{VJP over VJP (reverse over reverse)}
\end{align*}

A complete list of second-order terms related to the objective function $g$ and the three possible ways to compose the AD modes for each of the terms is shown in Tab.~\ref{Tab:g_related}.
\begin{table}[H]
\captionsetup{font=normalsize}
\caption{AD mode composition for $g$-related second-order terms}
\label{Tab:g_related}
\centering
\begin{spacing}{2}
\begin{tabular}{c|c|c|c|c}
\hline
AD mode composition
 &  \multicolumn{4}{c}{Second-order term} \\  \hline 
forward over reverse &   $\frac{\partial \Big( \frac{\partial g}{\partial \y} \Big)}{\partial \y} \hat{\y}$   &  $\frac{\partial \Big( \frac{\partial g}{\partial \y} \Big)}{\partial \thetab} \hat{\thetab}$    &   $\frac{\partial \Big( \frac{\partial g}{\partial \thetab} \Big)}{\partial \y} \hat{\y}$   & $\frac{\partial \Big( \frac{\partial g}{\partial \thetab} \Big)}{\partial \thetab} \hat{\thetab}$    \\
reverse over forward &  $\frac{\partial \Big( \frac{\partial g}{\partial \y} \hat{\y}\Big)}{\partial \y} $   &  $\frac{\partial \Big( \frac{\partial g}{\partial \thetab} \hat{\thetab}\Big)}{\partial \y} $    &  $\frac{\partial \Big( \frac{\partial g}{\partial \y} \hat{\y}\Big)}{\partial \thetab} $    & $\frac{\partial \Big( \frac{\partial g}{\partial \thetab} \hat{\thetab}\Big)}{\partial \thetab} $    \\
reverse over reverse & $\hat{\y}^T\frac{\partial \Big( \frac{\partial g}{\partial \y} \Big)}{\partial \y} $    &  $\hat{\thetab}^T\frac{\partial \Big( \frac{\partial g}{\partial \thetab} \Big)}{\partial \y} $   &  $\hat{\y}^T\frac{\partial \Big( \frac{\partial g}{\partial \y} \Big)}{\partial \thetab} $    &  $\hat{\thetab}^T\frac{\partial \Big( \frac{\partial g}{\partial \thetab} \Big)}{\partial \thetab} $   \\ \hline
\end{tabular}
\end{spacing}
\end{table}

A complete list of second-order terms related to the residual function $\rb$ and the three possible ways to compose the AD modes for each of the terms is shown in Tab.~\ref{Tab:r_related}.
\begin{table}[H]
\captionsetup{font=normalsize}
\caption{AD mode composition for $\rb$-related second-order terms}
\label{Tab:r_related}
\centering
\begin{spacing}{2}
\begin{tabular}{c|c|c|c|c}
\hline
AD mode composition
 &  \multicolumn{4}{c}{Second-order term} \\  \hline 
forward over reverse &   $\frac{\partial \Big( \lambdab^T\frac{\partial \rb}{\partial \y} \Big)}{\partial \y} \hat{\y}$   &  $\frac{\partial \Big( \lambdab^T\frac{\partial \rb}{\partial \y} \Big)}{\partial \thetab} \hat{\thetab}$    &   $\frac{\partial \Big( \lambdab^T\frac{\partial \rb}{\partial \thetab} \Big)}{\partial \y} \hat{\y}$   & $\frac{\partial \Big( \lambdab^T\frac{\partial \rb}{\partial \thetab} \Big)}{\partial \thetab} \hat{\thetab}$    \\
reverse over forward &  $\lambdab^T\frac{\partial \Big( \frac{\partial \rb}{\partial \y} \hat{\y}\Big)}{\partial \y} $   &  $\lambdab^T\frac{\partial \Big( \frac{\partial \rb}{\partial \thetab} \hat{\thetab}\Big)}{\partial \y} $    &  $\lambdab^T\frac{\partial \Big( \frac{\partial \rb}{\partial \y} \hat{\y}\Big)}{\partial \thetab} $    & $\lambdab^T\frac{\partial \Big( \frac{\partial \rb}{\partial \thetab} \hat{\thetab}\Big)}{\partial \thetab} $    \\
reverse over reverse & $\hat{\y}^T\frac{\partial \Big( \lambdab^T\frac{\partial \rb}{\partial \y} \Big)}{\partial \y} $    &  $\hat{\thetab}^T\frac{\partial \Big( \lambdab^T\frac{\partial \rb}{\partial \thetab} \Big)}{\partial \y} $   &  $\hat{\y}^T\frac{\partial \Big( \lambdab^T\frac{\partial \rb}{\partial \y} \Big)}{\partial \thetab} $    &  $\hat{\thetab}^T\frac{\partial \Big( \lambdab^T\frac{\partial \rb}{\partial \thetab} \Big)}{\partial \thetab} $   \\ \hline
\end{tabular}
\end{spacing}
\end{table}

We have implemented all three ways of AD mode composition for both $g$ and $\rb$ related second-order terms based on the fundamental JVP and VJP functions provided by \texttt{JAX} so that the implicit Hessian-vector product algorithm described in Alg.~(\ref{Alg:hessp}) can be completed.

\subsection{Accuracy}

To study the accuracy of the implemented implicit Hessian-vector product algorithm, we adopt two approaches: direct comparison with the finite difference method and the Taylor remainder test. 
The problem described in Section $\ref{Sec:model_problem}$ is used as a testbed.
This finite element problem is prescribed with the following conditions:
\begin{align}
\nonumber\Omega &=(0,1)\times(0,1) \quad \textrm{(A unit square)} \\
\nonumber b&=10\,\textrm{exp}\big(-((x_1-0.5)^2+(x_2-0.5)^2)/0.02 \big) \quad \textrm{(Source term)} \\
\nonumber\Gamma_D&=\{(0, x_2)\cup (1, x_2)\subset\partial\Omega\} \quad \textrm{(Dirichlet boundary)} \\
\nonumber\Gamma_N &=\{(x_1, 0)\cup (x_1, 1)\subset\partial\Omega\} \quad \textrm{(Neumann boundary)} \\
\nonumber u_D&=0 \quad \textrm{(Dirichlet boundary condition)} \\
t&=\textrm{sin}(5x_1) \quad \textrm{(Neumann boundary condition)}
\end{align}

The problem is discretized using 4096 linear quadrilateral elements in 2D.
The parameter vector $\thetab=[\theta_1,\theta_2,\dots,\theta_M]$ is the discretized version of the parameter field $\theta(\x)$, where $M=4096\times 4$ due to the 4 quadrature points used for integration of each element.
The output solution vector $\y=[u_1, u_2,\dots,u_N]$ is the discretized version of $u(\x)$, and $N=4225$ is the total number of degrees of freedom.

\begin{figure}[H] 
\centering
\includegraphics[scale=0.5]{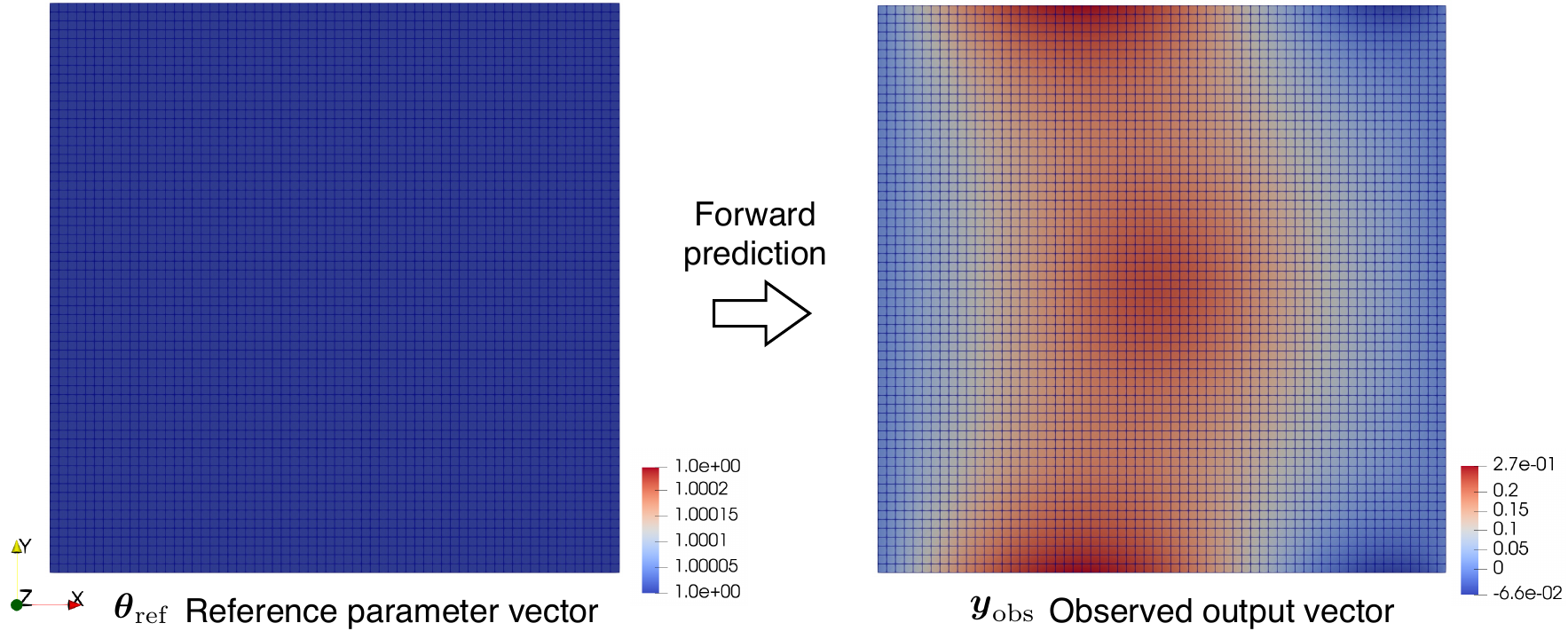}
\caption{The forward prediction part of the model problem for implicit differentiation.}
\label{Fig:model_problem}
\end{figure}

As shown in Fig.~\ref{Fig:model_problem}, a reference parameter vector $\thetab_{\textrm{ref}}$ (all entries set to be one) is used to parametrize the physical system, and the forward problem is solved using the finite element method to generate the corresponding output solution vector $\y_{\textrm{obs}}$.
This output $\y_{\textrm{obs}}$ is stored as the known observation data, i.e., (discretized) $u_{\textrm{obs}}$ in Eq.~(\ref{Eq:PDE-CO}).
To solve the inverse optimization problem in Eq.~(\ref{Eq:PDE-CO_discretized}), derivative information is critical.
In particular, we will study the accuracy of the Hessian-vector product, i.e., $\frac{\textrm{d}^2 g}{\textrm{d}\thetab^2} \hat{\thetab}$ (or $\Hb\hat{\thetab}$).

\subsubsection{Comparison with finite difference approximation}

Analytical Hessian-vector product is usually not available.
One way to check the accuracy of the Hessian-vector product computed by the proposed implicit differentiation method is to directly compare the results with the finite difference method.

We use the central difference method to approximate the Hessian-vector product as
\begin{align} \label{Eq:FD}
    \vb_{\textrm{FD}}  = \Big( \frac{\textrm{d} g}{\textrm{d}\thetab}\big(\thetab + h \hat{\thetab} \big) - \frac{\textrm{d} g}{\textrm{d}\thetab }\big(\thetab - h \hat{\thetab} \big) \Big)/\big(2h\big),
\end{align}
where $\frac{\textrm{d} g}{\textrm{d}\thetab}$ is the gradient function defined by the first-order implicit differentiation method~\cite{xue2023jax}.

We define the Hessian-vector product evaluated by the proposed method as 
\begin{align}
    \vb_{\textrm{AD}} = \Hb\big(\thetab\big) \hat{\thetab},
\end{align}
where we anticipate that $\vb_{\textrm{AD}}$ gives exact Hessian-vector product up to machine precision, and that $\vb_{\textrm{FD}}\approx\vb_{\textrm{AD}}$.

Therefore, the relative difference between our implicit differentiation approach and the finite difference approach is
\begin{align} \label{Eq:error_v}
    e_{\vb} = \frac{\Vert \vb_{\textrm{AD}}- \vb_{\textrm{FD}}\Vert}{\Vert \vb_{\textrm{AD}}\Vert}.
\end{align}

\begin{figure}[H] 
\centering
\includegraphics[scale=0.6]{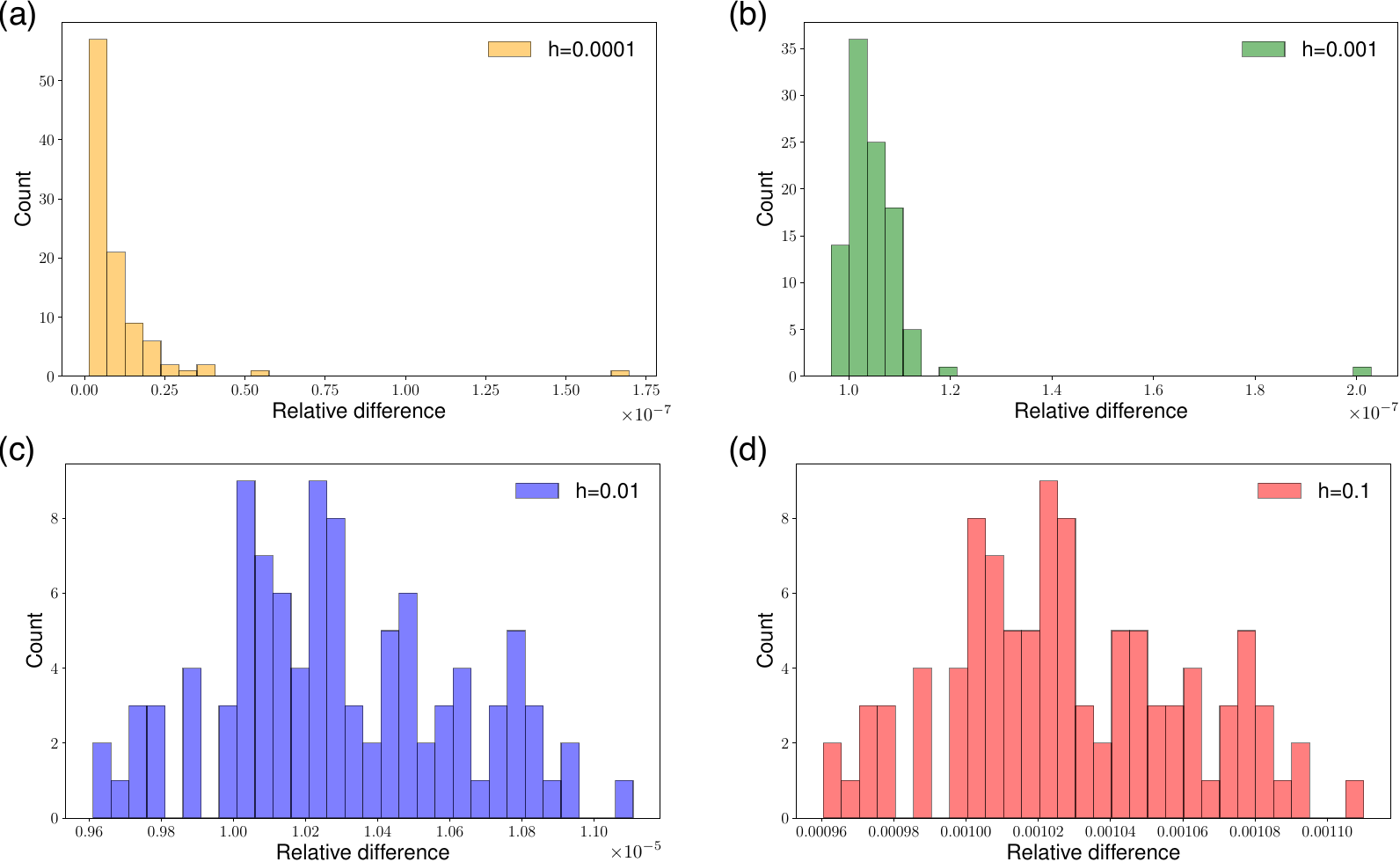}
\caption{Relative difference $e_{\vb}$ of Hessian-vector product evaluations between the proposed implicit differentiation approach and the finite difference approach. In each subfigure, 100 samples are used to generate the histogram plot.}
\label{Fig:Hv_fwd_rev}
\end{figure}

We set the step size $h$ to be $h=10^{-4}, 10^{-3}, 10^{-2}, 10^{-1}$.
For each step size $h$, we generate 100 standard normal random vectors (all entries are independent and each entry follows the standard normal distribution) of size $M$ for the parameter vector $\thetab$ and another 100 standard normal random vectors of size $M$ for the incremental vector $\hat{\thetab}$.
Based on each pair $(\thetab, \hat{\thetab})$, $\vb_\textrm{FD}$ and $\vb_\textrm{AD}$ are evaluated to obtain the relative difference $e_{\vb}$.
The histogram plots for $e_{\vb}$ are shown in Fig.~\ref{Fig:Hv_fwd_rev} for all the four step sizes.

As shown in Fig.~\ref{Fig:Hv_fwd_rev}, the relative difference between the implicit differentiation approach and the finite difference approach is small in general, with the maximum difference to occur around $0.110\% \sim 0.112\%$ for $h=0.1$.
The overall trend of the relative difference decreases as the step size $h$ decrease, meaning that the finite difference results are closer to the implicit differentiation results with smaller step sizes.
Therefore, the results agree with our expectation that the implicit differentiation approach gives accurate Hessian-vector products.

Similar to the vector-based relative difference defined in Eq.~(\ref{Eq:error_v}), we consider scalar-based relative difference
\begin{align} \label{Eq:error_s}
    e_{s} = \frac{|s_\textrm{FD} - s_\textrm{AD}|}{|s_\textrm{AD}|},
\end{align}
where with an applied vector $\tilde{\thetab}$, the vector-Hessian-vector product based on finite difference approach $s_\textrm{FD}$ is defined as 
\begin{align}
    s_\textrm{FD} = \tilde{\thetab}^T \vb_\textrm{FD},
\end{align}
and the vector-Hessian-vector product based on implicit differentiation approach $s_\textrm{AD}$ is defined as 
\begin{align}
    s_\textrm{AD} = \tilde{\thetab}^T \vb_\textrm{AD}.
\end{align}

\begin{figure}[H] 
\centering
\includegraphics[scale=0.6]{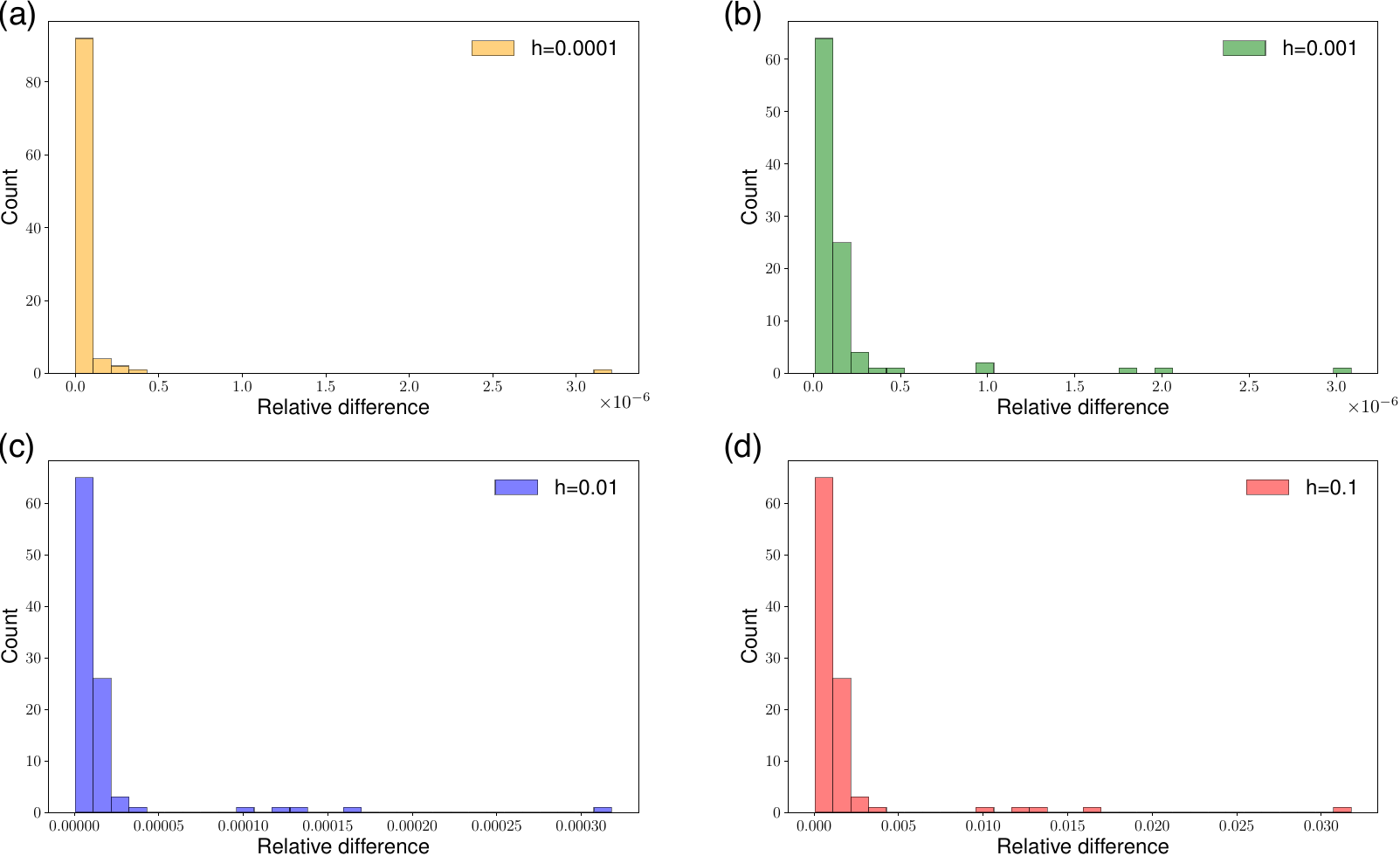}
\caption{Relative difference $e_{s}$ of Hessian-vector product evaluations between the proposed implicit differentiation approach and the finite difference approach. In each subfigure, 100 samples are used to generate the histogram plot.}
\label{Fig:vHv_fwd_rev}
\end{figure}

Here, for each step size $h$, we generate 100 standard normal random vectors of size $M$ each for the parameter vector $\thetab$, the right incremental vector $\hat{\thetab}$, and the left incremental vector $\tilde{\thetab}$.
Based on each triple $(\thetab, \hat{\thetab}, \tilde{\thetab})$, $s_\textrm{FD}$ and $s_\textrm{AD}$ are evaluated to get the relative difference $e_{s}$.
The histogram plots for $e_{s}$ are shown in Fig.~\ref{Fig:vHv_fwd_rev} for all the four step sizes.

As shown in Fig.~\ref{Fig:vHv_fwd_rev}, the largest relative difference between the implicit differentiation approach and the finite difference approach is around $3\%$ for $h=0.1$.
The overall trend of the relative difference decreases as the step size $h$ decrease, meaning that the finite difference results are closer to the implicit differentiation results with smaller step sizes.
Therefore, these results of scalar-based relative difference agree with our expectation that the implicit differentiation approach is accurate in evaluating Hessian-vector products.

The results of implicit differentiation in Fig.~\ref{Fig:Hv_fwd_rev} and Fig.~\ref{Fig:vHv_fwd_rev} above adopt the ``forward over reverse'' AD mode composition.
We have also considered the other two ways of AD mode composition (``reverse over forward'' and ``reverse over reverse''), and the results are similar; hence, not shown here.
The three ways of AD mode composition all give accurate Hessian-vector products, and the maximum mutual relative difference between them is smaller than $10^{-12}$, measured by the scalar-based metric in Eq.~(\ref{Eq:error_s}).

\subsubsection{Taylor remainder test}

As suggested by~\cite{mitusch2019dolfin}, the Taylor remainder test is a useful tool to verify the gradients, which has been used in several previous works~\cite{niewiarowski2020adjoint,xue2022mapped}.
Since we are concerned with Hessians in this work, the Taylor remainder test is extended to be used in verification of Hessians.

According to the Taylor's theorem~\cite{MR0055409}, the convergence rate of the residual should be 1 using a zeroth-order expansion:
\begin{align} \label{Eq:zeroth_expansion}
    r_{\textrm{zeroth}}=\big| g(\thetab + \epsilon\delta\thetab) - g(\thetab) \big| \rightarrow 0 \textrm{ at } \mathcal{O}(\epsilon),
\end{align}
where $\delta \thetab \in \mathbb{R}^M$ is a given direction of perturbation, $\epsilon\in\mathbb{R}$ is the scaling factor, and $g$ is the objective function under consideration. 
The convergence rate of the residual should be 2 by a first-order expansion:
\begin{align}\label{Eq:first_expansion}
    r_{\textrm{first}}=\big| g(\thetab + \epsilon\delta\thetab) - g(\thetab) - \epsilon \,\frac{\textrm{d}g}{\textrm{d}\,\thetab}\delta \thetab \big| \rightarrow 0 \textrm{ at } \mathcal{O}(\epsilon^2).
\end{align}
Similarly, the convergence rate should be 3 by a second-order expansion:
\begin{align}\label{Eq:second_expansion}
    r_{\textrm{second}}=\big| g(\thetab + \epsilon\delta\thetab) - g(\thetab) - \epsilon \,\frac{\textrm{d}g}{\textrm{d}\thetab}\,\delta \thetab  - \epsilon^2 \,\delta \thetab^T  \,\frac{\textrm{d}^2 g}{\textrm{d}\thetab^2} \, \delta\thetab \big| \rightarrow 0 \textrm{ at } \mathcal{O}(\epsilon^3).
\end{align}

\begin{figure}[H] 
\centering
\includegraphics[scale=0.45]{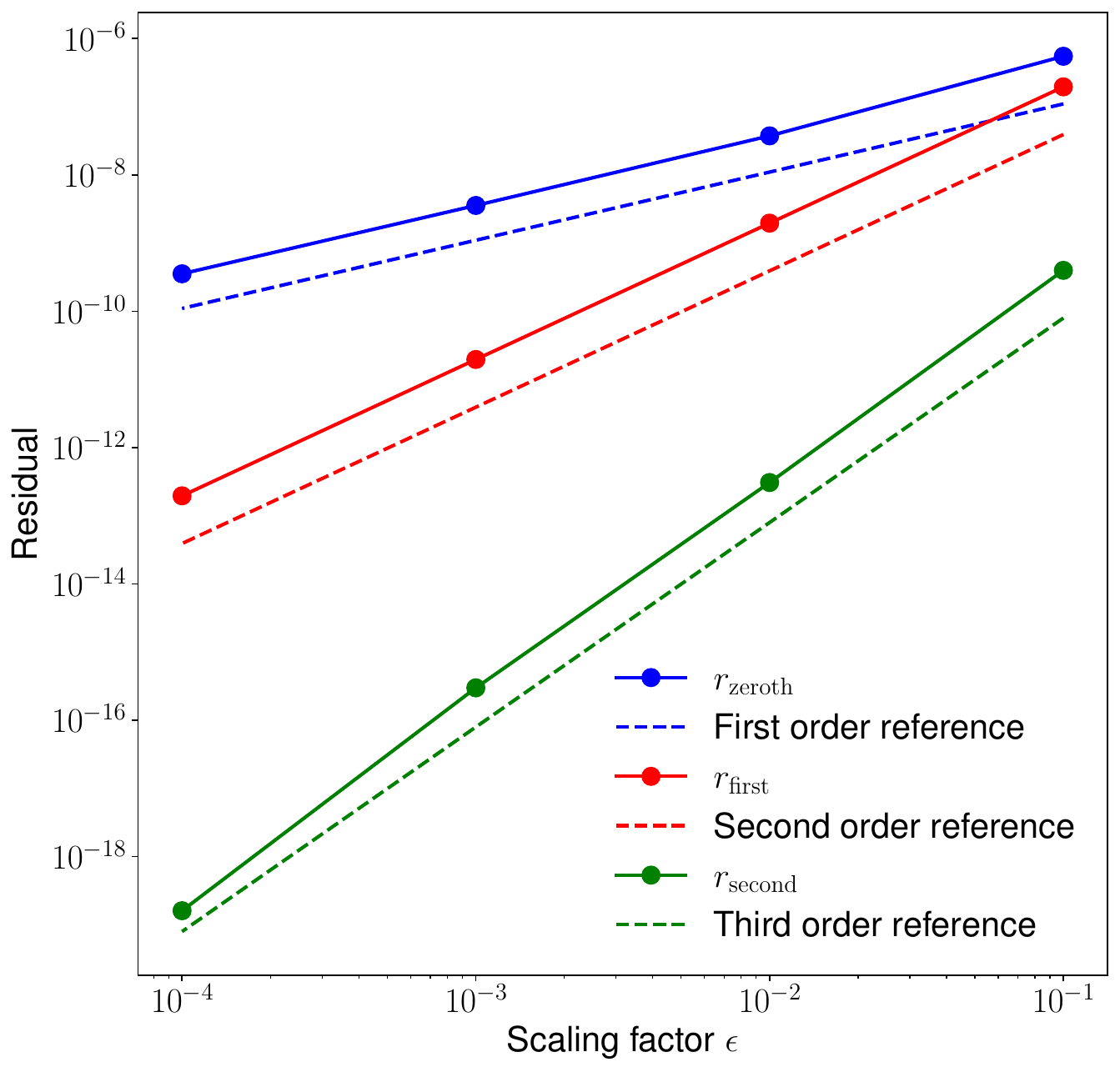}
\caption{Taylor remainder test. As expected, the zeroth-order expansion of the residual achieves a first order convergence, the first-order expansion achieves a second order convergence, and the second-order expansion achieves a third order convergence.}
\label{Fig:Taylor}
\end{figure}

We set the scaling factor $\epsilon$ to be $\epsilon=10^{-4}, 10^{-3}, 10^{-2}, 10^{-1}$ and compute the residuals $r_{\textrm{zeroth}}$, $r_{\textrm{first}}$ and $r_{\textrm{second}}$ under these scaling factors.
The results are shown in Fig.~\ref{Fig:Taylor}.
As expected, we achieve $r_{\textrm{zeroth}} \propto \epsilon$, $r_{\textrm{first}} \propto \epsilon^2$ and $r_{\textrm{second}} \propto \epsilon^3$, showing the correct computation of the Hessian-vector product.

\subsection{Efficiency}

To compute Hessian-vector products, the finite difference approach in Eq.~(\ref{Eq:FD}) needs to evaluate the gradient $\frac{\textrm{d}g}{\textrm{d}\thetab}$ at two different locations $\thetab+h\hat{\thetab}$ and $\thetab-h\hat{\thetab}$.
For each gradient evaluation, the forward problem and the adjoint problem must be solved.
The proposed implicit differentiation approach must solve the forward problem, the adjoint problem, the incremental forward problem and the incremental adjoint problem.
Loosely speaking, both approaches need to solve four problems to obtain a Hessian-vector product, so the computational cost is roughly on the same scale. 

The implicit differentiation approach might have some advantages in two situations.
The first situation is if the forward problem is nonlinear and requires significant computational cost. 
Then the implicit differentiation approach wins in that it only needs to solve the forward problem once, while the finite difference approach needs to solve it twice.
Note that the adjoint problem, the incremental forward problem, and the incremental adjoint problem are always linear.
The second situation is that when $\thetab$ is fixed and $\hat{\thetab}$ needs to be frequently updated.
Then the implicit differentiation approach can cache the  solution variable $\y$ and the adjoint variable $\lambdab$, and only needs to solve the incremental forward and adjoint problems for any updated $\hat{\thetab}$ with the same $\thetab$.
The finite difference approach cannot avoid solving the forward problems.

\begin{figure}[H] 
\centering
\includegraphics[scale=0.6]{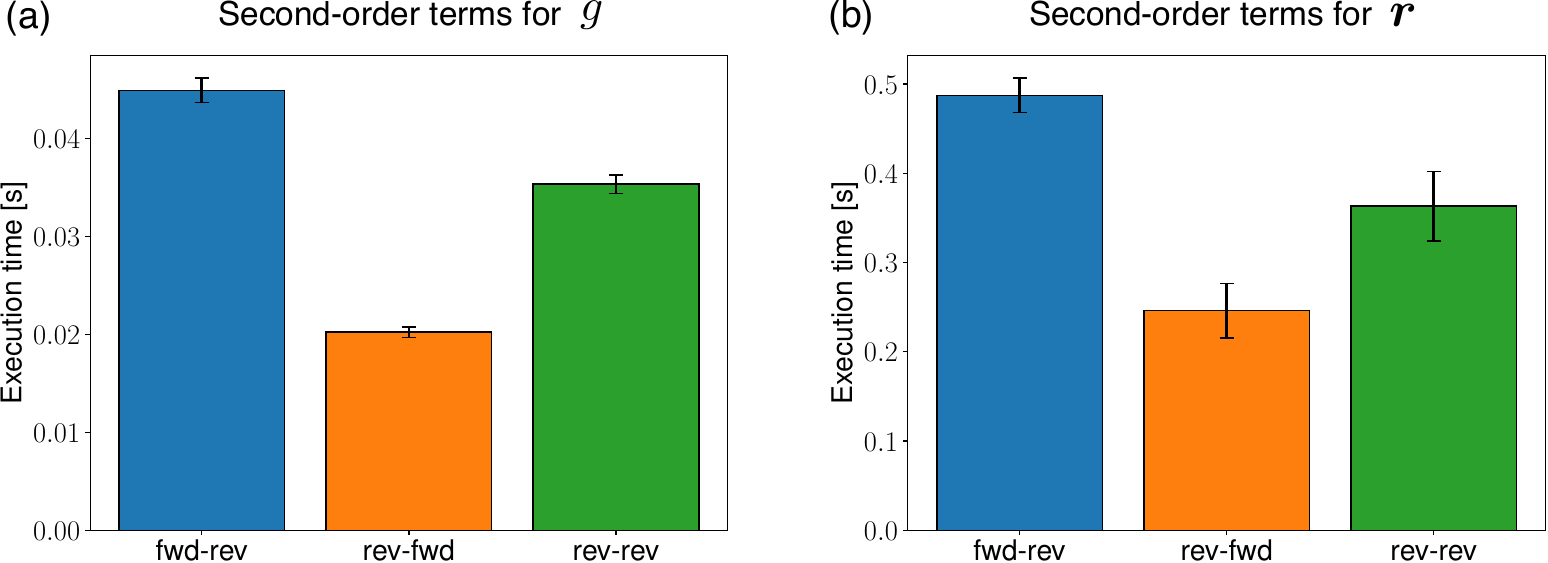}
\caption{Execution time for the three ways of AD mode composition: ``fwd-rev'' stands for ``forward over reverse'', ``rev-fwd'' stands for ``reverse over forward'', and ``rev-rev'' stands for ``reverse over reverse''.  (a) The time for evaluating the objective function $g$-related second order terms; (b) The time for evaluating the residual function $\rb$-related second order terms.}
\label{Fig:profiling}
\end{figure}

Despite the advantages mentioned above, the implicit differentiation approach must compute the second-order terms as listed in Tab.~\ref{Tab:g_related} and Tab.~\ref{Tab:r_related} using either way of the AD mode composition from the three choices. 
To investigate the computational cost of these second-order terms, we profile the three ways of AD mode composition for both $g$-related and $\rb$-related terms.
Each execution is repeated 11 times and the first execution is dropped due to the overhead time caused by \texttt{JAX}'s tracing and compilation process.  
The profiling report is shown in Fig.~\ref{Fig:profiling}.
The most efficient mode of composition is the reverse mode over the forward mode, which is VJP over JVP.
Therefore, we will stick to this mode in the following discussions of numerical experiments. 

A potential direction to accelerate the evaluations of these second-order derivative terms is to utilize sparse properties of the underlying system.
In the current implementation, only the sparsity of $\frac{\partial \rb}{\partial \y}$ is explicitly considered through the finite element sparse matrix assembly following our previous work \texttt{JAX-FEM}.
The sparsity of second-order terms might be exploited to accelerate the computation.
For future endeavors, we point to automatic sparse differentiation (ASD), a promising area under active research~\cite{hill2025sparser}.
 
\section{Numerical examples}
\label{Sec:numerical}

To solve PDE-constrained optimization problems, we explore two popular optimization algorithms: the L-BFGS-B method~\cite{zhu1997algorithm} and the Newton-CG method~\cite{fletcher1964function,nash1984newton}.
Both methods depend on the information of derivatives.
However, the L-BFGS-B method avoids explicit Hessian-vector products by approximating curvature through gradient history, while the Newton-CG method directly relies on Hessian-vector products to solve the Newton equation iteratively.
For the Newton-CG method, Hessian-vector products can be found by the approximate finite difference (FD) approach or the proposed implicit automatic differentiation (AD) approach.
Therefore, we have three options for optimization: L-BFGS-B, Newton-CG (FD), and Newton-CG (AD).
The \texttt{SciPy}~\cite{virtanen2020scipy} package is employed for the L-BFGS-B method and the Newton-CG method.

As shown in Tab.~\ref{Tab:specs}, the numerical examples of PDE-constrained optimization problems to be presented cover a wide range of tasks with diverse possibilities of optimization parameters (weak form terms and boundary conditions).
The corresponding forward problems can be either linear or nonlinear, 2D or 3D, and the unknown solutions range from a single variable to coupled variables.
The computational platform is a personal MacBook Pro laptop of year 2019 with a processor of 2.4 GHz 8-Core Intel Core i9, which is sufficient to solve most of the following inverse optimization problems within one minute.

\begin{table}[H]
\captionsetup{font=normalsize}
\caption{General specifications of the four numerical examples}
\label{Tab:specs}
\centering
\begin{tabular}{lllll}
\hline
Task                          & Optimization parameters      & Linearity & Dimension & Variable \\ \hline
Source field identification   & Source term                  & Linear    & 2D        & Single   \\
Traction force identification & Neumann boundary condition   & Nonlinear & 3D        & Single   \\
Thermal-mechanical control    & Dirichlet boundary condition & Linear    & 2D        & Coupled  \\
Shape optimization            & Stiffness term               & Nonlinear & 2D        & Single   \\ \hline
\end{tabular}
\end{table}

\subsection{Source field identification}
 
We consider an inverse problem of source field identification based on the observed solution field. 
The observed data is generated by solving a forward linear Poisson's problem, whose governing equation states that find $u: \Omega \rightarrow \mathbb{R}$ such that
\begin{align} \label{Eq:strong_poisson}
    -\nabla \cdot \big( \nabla u \big) = b & \quad \textrm{in}  \, \, \Omega, \nonumber \\
    u = 0 &  \quad\textrm{on} \, \, \Gamma_D,  \nonumber \\
  \nabla u \cdot \nb = 0  & \quad \textrm{on} \, \, \Gamma_N.
\end{align}
where $\Omega\subset\mathbb{R}^{\textrm{dim}}$ is the problem domain, $b:\Omega\rightarrow\sR$ is the source field (parameter field).

The weak form of Eq.~(\ref{Eq:strong_poisson}) states the following: Find the solution $u$ such that for any test function $v$
\begin{align} \label{Eq:weak_poisson}
  r(u; v) = \int_{\Omega}  \nabla u  \cdot  \nabla v \textrm{ d} \Omega - \int_{\Omega} b \,  v \textrm{ d} \Omega = 0.
\end{align}

This finite element problem is prescribed with the following conditions:
\begin{align}
\nonumber\Omega &=(0,1)\times(0,1) \quad \textrm{(A unit square)} \\
\nonumber\Gamma_D&=\{(0, x_2)\cup (1, x_2)\subset\partial\Omega\} \quad \textrm{(Dirichlet boundary)} \\
\nonumber\Gamma_N &=\{(x_1, 0)\cup (x_1, 1)\subset\partial\Omega\} \quad \textrm{(Neumann boundary)}
\end{align}

The problem is discretized using 4096 linear quadrilateral elements in 2D.
The parameter vector $\thetab =[b_1,b_2,\dots,b_M]$ is the discretized version of the parameter field $b(\x)$, where $M=4096\times 4$ due to the 4 quadrature points used for the integration of each element.
The output solution vector $\y=[u_1, u_2,\dots,u_N]$ is the discretized version of $u(\x)$, where $N=4225$ is the total number of degrees of freedom.

\begin{figure}[H] 
\centering
\includegraphics[scale=0.48]{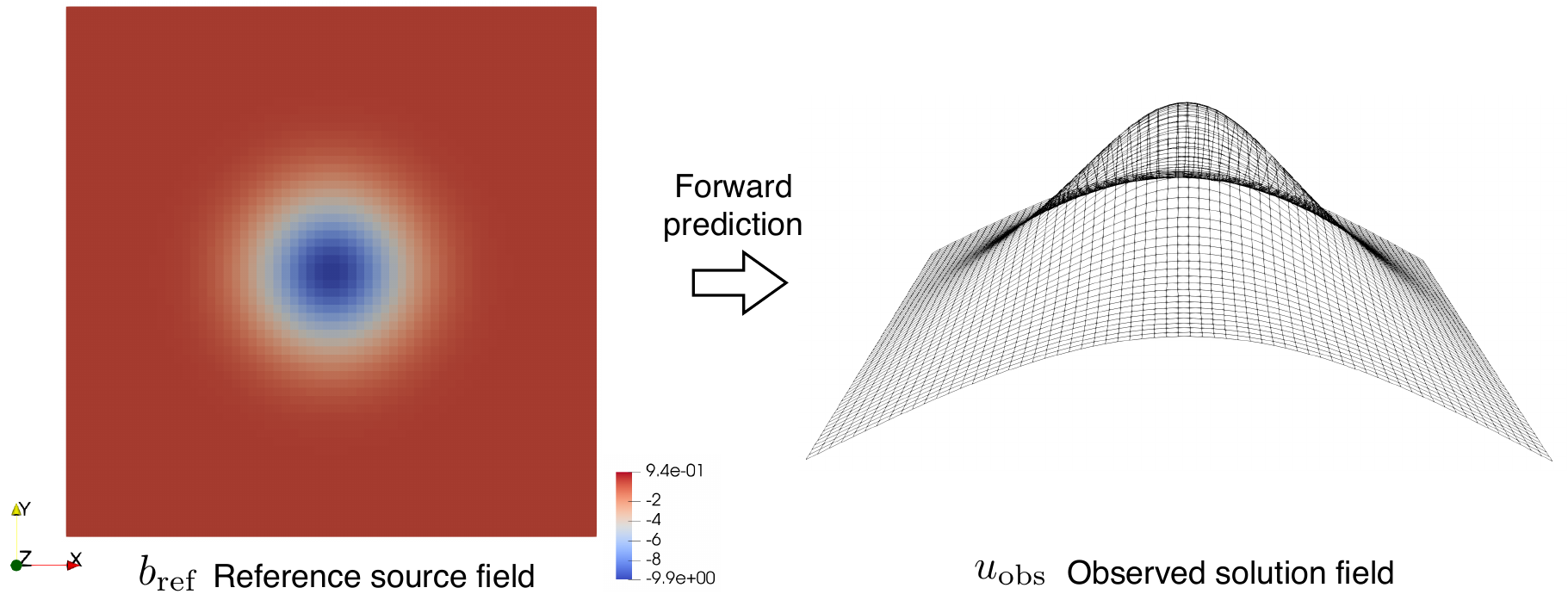}
\caption{The forward prediction of the linear Poisson's problem. }
\label{Fig:poisson_fwd}
\end{figure}

To generate the observed data $u_{\textrm{obs}}(\x)$, we set a reference source field as $b_{\textrm{ref}}(\x)=10\,\textrm{exp}\big(-((x_1-0.5)^2+(x_2-0.5)^2)/0.02 \big)$.
Based on this reference source field $b_{\textrm{ref}}$, we solve the forward prediction problem and obtain $u_{\textrm{obs}}$.
Fig.~\ref{Fig:poisson_fwd} shows the reference source field $b_{\textrm{ref}}$ and the corresponding solution field $u_{\textrm{obs}}$.

Based on the observed data $u_{\textrm{obs}}$, the inverse parameter identification problem is defined as a PDE-constrained optimization problem:
\begin{align} \label{Eq:PDE-CO_poisson}
    \nonumber &\min_{u,b}  \frac{1}{2}\int_{\Omega}  (u - u_{\textrm{obs}})^2 \textrm{ d} \Omega
        + \frac{\alpha}{2} \int_{\Omega} b^2 \textrm{ d} \Omega \\
    & \qquad \qquad \textrm{s.t.} \quad r(u;v)=0, 
\end{align}
where the source field $b$ is optimized to encourage the predicted solution field $u$ to match the observed solution field $u_{\textrm{obs}}$ with a regularization term.

\begin{figure}[H] 
\centering
\includegraphics[scale=0.45]{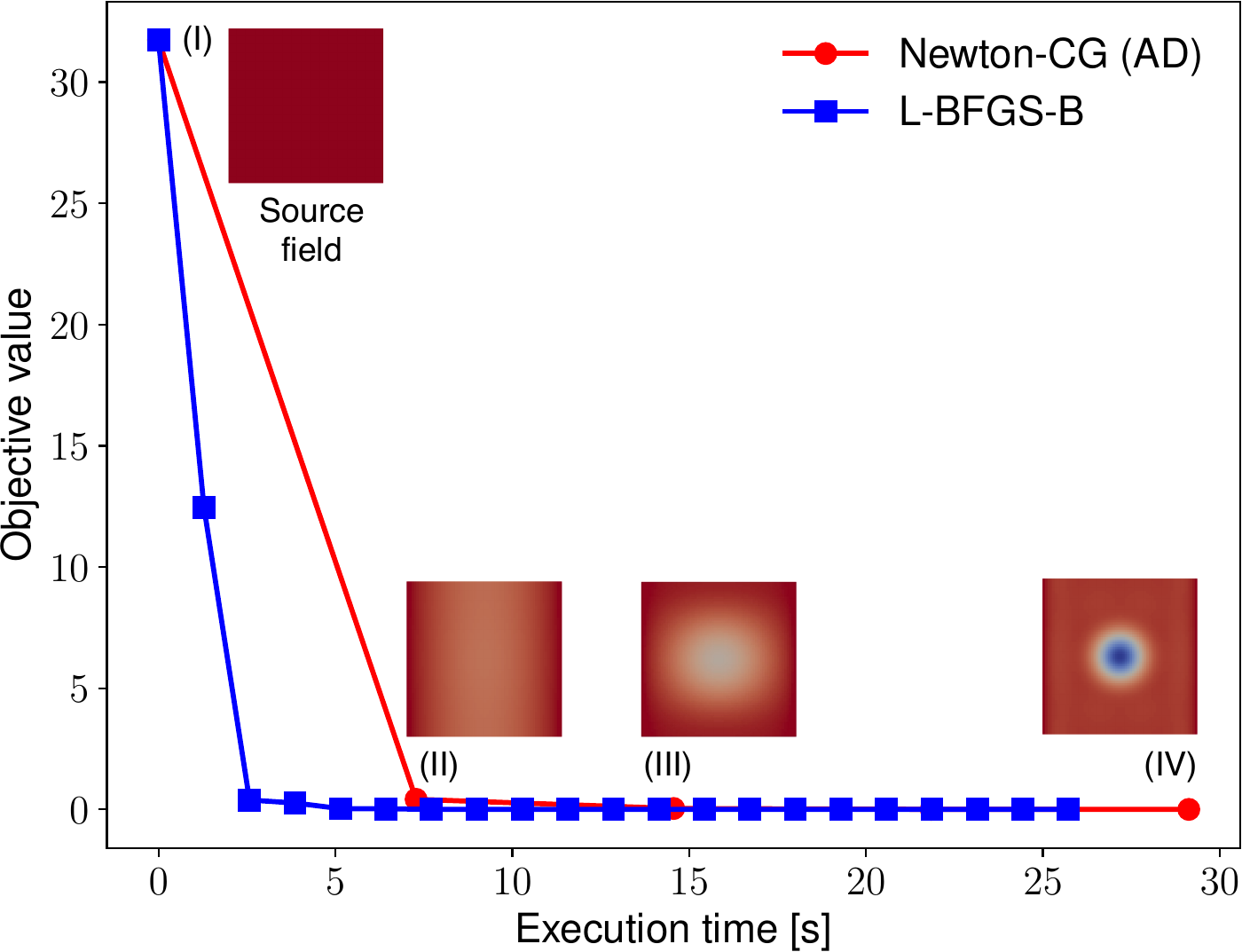}
\caption{Optimization results for the source field identification problem. The source fields at iteration 0, 1, 2 and 4 of Newton-CG (AD) are marked as (I), (II), (III) and (IV). }
\label{Fig:poisson_opt}
\end{figure}

The three optimizers L-BFGS-B, Newton-CG (FD) and Newton-CG (AD) are used to solve the PDF-constrained optimization problem, and the results are shown in Fig.~\ref{Fig:poisson_opt}, where the objective value is plotted against the actual execution time.
Each data point in Fig.~\ref{Fig:poisson_opt} is recorded at every optimization iteration (defined by the callback function of \texttt{SciPy}) for both L-BFGS and Newton-CG (AD). 
Note that the result of Newton-CG (FD) is missing, since the optimizer does not converge and cannot give a valid result.
In contrast, Newton-CG (AD) converges and the optimization is successful.
This contrast between Newton-CG (AD) and Newton-CG (FD) indicates the importance of accurate Hessian-vector products instead of approximate ones.

\begin{figure}[H] 
\centering
\includegraphics[scale=0.65]{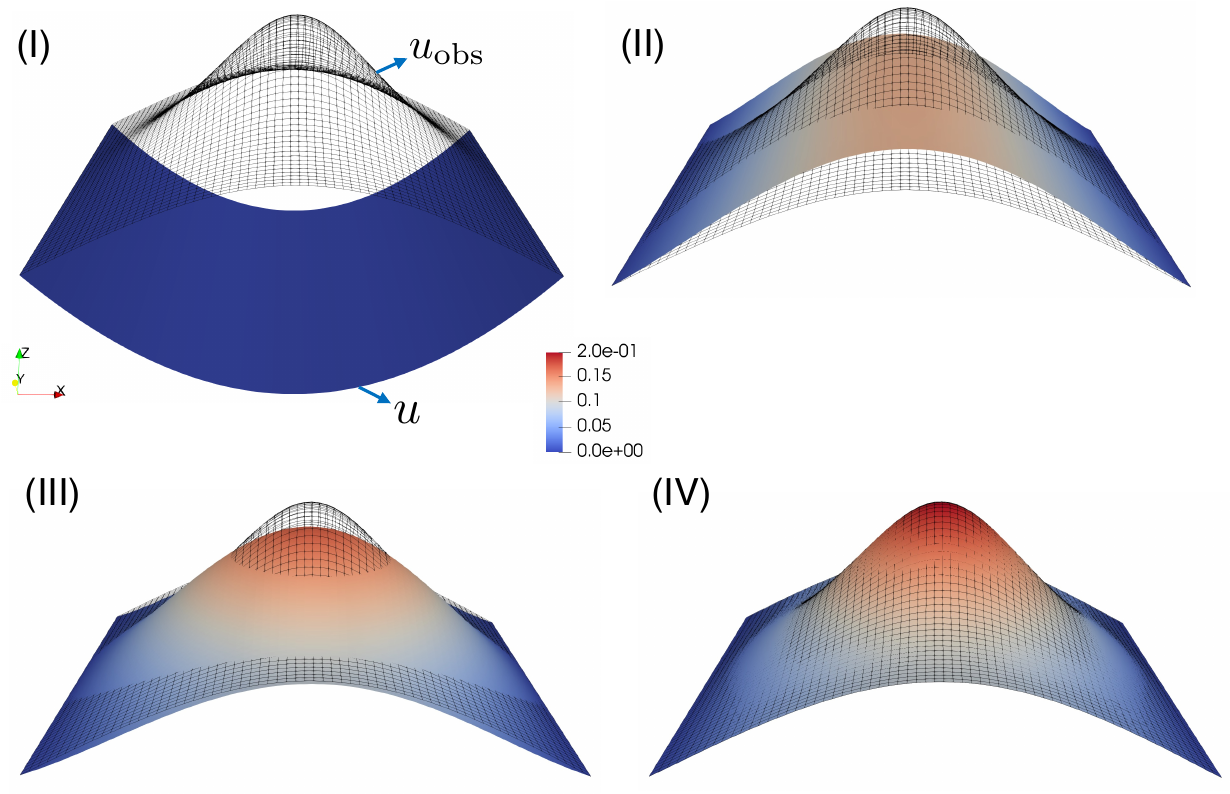}
\caption{Matching between the predicted solution fields (colored plot) and the observed solution field $u_\textrm{obs}$ (wireframe plot) at optimization steps (I) to (IV).}
\label{Fig:poisson_inv}
\end{figure}

In Fig.~\ref{Fig:poisson_opt}, we mark four representative iteration steps of Newton-CG (AD) from step (I) to step (IV), and show the source fields at those steps.
We see that the final source field at step (IV) is close to the reference source field $b_\textrm{ref}$ in Fig.~\ref{Fig:poisson_fwd}, and the objective value is close to zero, showing that the source field identification task is successful.
To further show that the predicted solution field indeed matches the observed solution field $u_{\textrm{obs}}$, we show the predicted solution fields at step (I) to (IV) in Fig.~\ref{Fig:poisson_inv}, along with the observed solution field $u_\textrm{obs}$.
As shown, the predicted solution field at the final step (IV) matches well with the observed solution field $u_\textrm{obs}$.

In this numerical example, we conclude that both L-BFGS-B and Newton-CG (AD) work reasonably well, and L-BFGS-B is recommended due to its faster convergence to a low objective value.
Newton-CG (FD) fails to generate any useful results.

\subsection{Traction force identification}
\label{Sec:traction}

We consider a task of identifying the boundary traction force for a hyperelastic material based on an observed displacement field, a similar task recently studied by~\cite{mei2024non}.
Traction force identifications are practically useful in various fields such as aerospace or biomedical engineering.
Here, we consider a typical neo-Hookean solid, whose governing equation of static equilibrium is 
\begin{align} \label{Eq:strong_hyperelasticity}
    -\nabla \cdot \Pb = \boldsymbol{0} & \quad \textrm{in}  \, \, \Omega, \nonumber \\
    \ub = \boldsymbol{0} &  \quad\textrm{on} \, \, \Gamma_D,  \nonumber \\
    \Pb \cdot \nb = \tb  & \quad \textrm{on} \, \, \Gamma_N,
\end{align}
where $\Pb$ is the first Piola-Kirchhoff stress defined through the strain energy density function $W$
\begin{align} \label{Eq:neo-hookean}
    \Pb &= \frac{\partial W}{\partial \Fb}, \nonumber \\
    \Fb &= \nabla \ub + \Ib, \nonumber \\
    W (\Fb) &= \frac{G}{2}(J^{-2/3} I_1 - 3) + \frac{\kappa}{2}(J - 1)^2,
\end{align}
where $\ub:\Omega \rightarrow \mathbb{R}^3$ is the displacement field to be solved, $\Fb$ is the deformation gradient, $J = \textrm{det}(\Fb)$,~$I_1 = \textrm{tr}(\Fb^T\Fb)$;~$G = \frac{E}{2(1+\nu)}$ and~$\kappa = \frac{E}{3(1-2\nu)}$ denote the initial shear and bulk moduli, respectively, with~$E$ being the Young's modulus and and~$\nu$ the Poisson's ratio of the material.
The above~${W}$ is commonly used to model isotropic elastomers that are almost incompressible~\cite{ogden1997non}.

The weak form of Eq.~(\ref{Eq:strong_hyperelasticity}) states the following: Find the solution $\ub$ such that for any test function $\vb$
\begin{align} \label{Eq:weak_hyperelasticity}
    r(\ub; \vb) = \int_{\Omega}  \boldsymbol{P} : \nabla \boldsymbol{v} \, \textrm{d}\Omega - \int_{\Gamma_N} \boldsymbol{t} \cdot \boldsymbol{v} \, \textrm{d}\Gamma = 0.
\end{align}

This finite element problem is prescribed with the following conditions:
\begin{align}
\nonumber\Omega &=(0,1)\times(0,1) \times (0,0.05) \quad \textrm{(A thin plate)} \\
\nonumber\Gamma_D&=\{(x_1, 0, x_3)\subset\partial\Omega\} \quad \textrm{(Dirichlet boundary)} \\
\nonumber\Gamma_N &=\{(x_1, 1, x_3)\subset\partial\Omega\} \quad \textrm{(Neumann boundary, non-zero traction)}
\end{align}

\begin{figure}[H] 
\centering
\includegraphics[scale=0.55]{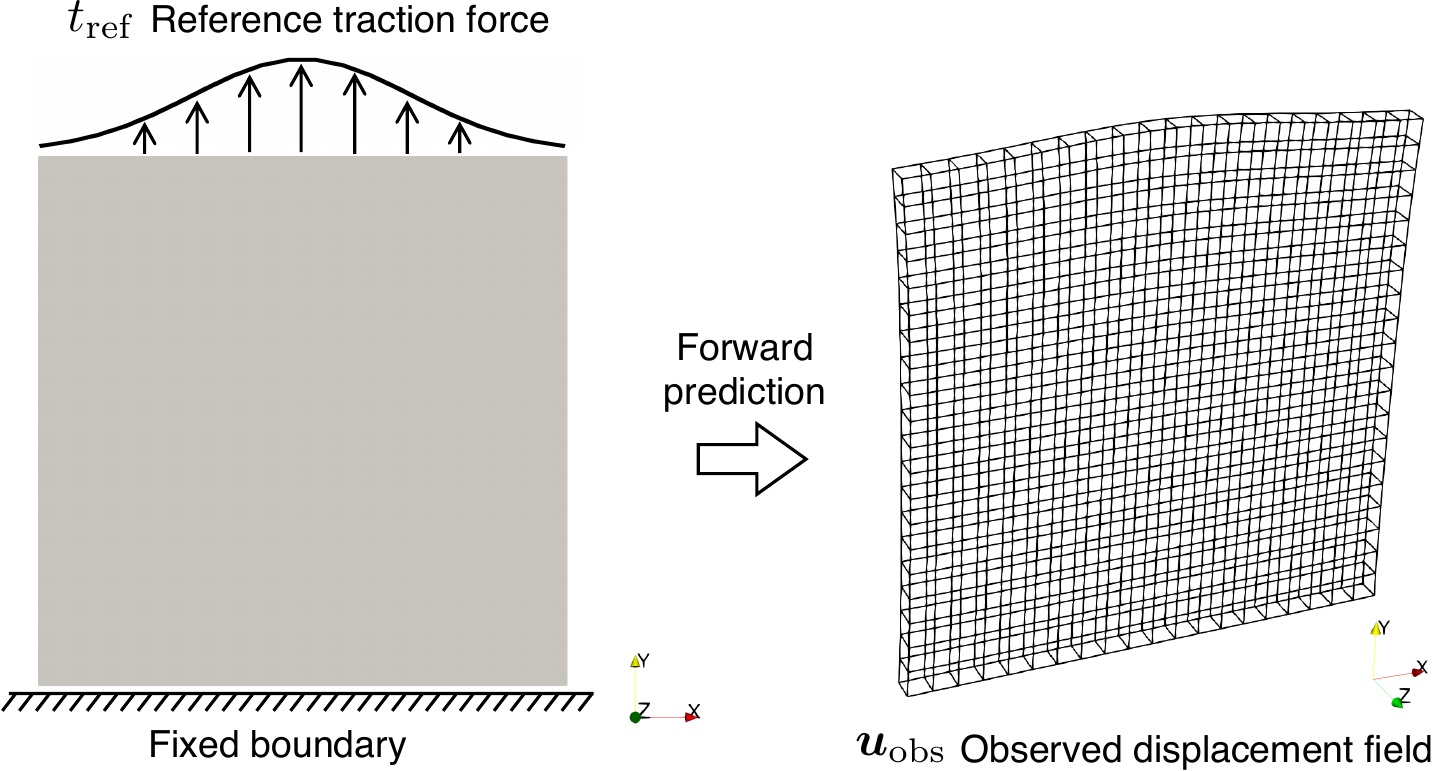}
\caption{The forward prediction of the hyperelasticity problem. }
\label{Fig:hyperelasticity_fwd}
\end{figure}

The problem is discretized using 400 linear hexahedron elements in 3D.
The parameter field is the traction force $t(\x)$ along the $x_2$-axis on the top side of the thin plate (see the left panel of Fig.~\ref{Fig:hyperelasticity_fwd}.
The parameter vector $\thetab =[t_1,t_2,\dots,t_M]$ is the discretized version of the parameter field $t(\x)$, where $M=20\times 4$ due to the 20 element faces on the top side and the 4 quadrature points used for the surface integration of each element face.
The output solution vector $\y=[u_1, u_2,\dots,u_N]$ is the discretized version of $\ub(\x)$, where $N=882\times 3$ due to the 882 finite element nodes and the 3 degrees of freedom associated with each node.

To generate the observed data $\ub_{\textrm{obs}}(\x)$, we set a reference traction force field as $t_{\textrm{ref}}(\x)=10^5\,\textrm{exp}\big(-(x_1-0.5)^2/0.08 \big)$.
Based on this reference traction force $t_{\textrm{ref}}$, we solve the forward prediction problem and obtain $\ub_{\textrm{obs}}$.
Fig.~\ref{Fig:hyperelasticity_fwd} shows the reference traction force field $t_{\textrm{ref}}$ and the corresponding solution field $\ub_{\textrm{obs}}$.

Based on the observed data $\ub_{\textrm{obs}}$, the inverse parameter identification problem is defined as a PDE-constrained optimization problem:
\begin{align} \label{Eq:PDE-CO_hyperelasticity}
    \nonumber &\min_{\ub,t}  \frac{1}{2}\int_{\Omega}  \Vert\ub - \ub_{\textrm{obs}}\Vert^2 \textrm{ d} \Omega
        + \frac{\alpha}{2} \int_{\Gamma_N} t^2 \textrm{ d} \Gamma \\
    & \qquad \qquad \textrm{s.t.} \quad r(\ub;\vb)=0, 
\end{align}
where the traction force field $t$ is optimized to encourage the predicted displacement field $\ub$ to match the observed displacement field $\ub_{\textrm{obs}}$ with a regularization term.

\begin{figure}[H] 
\centering
\includegraphics[scale=0.5]{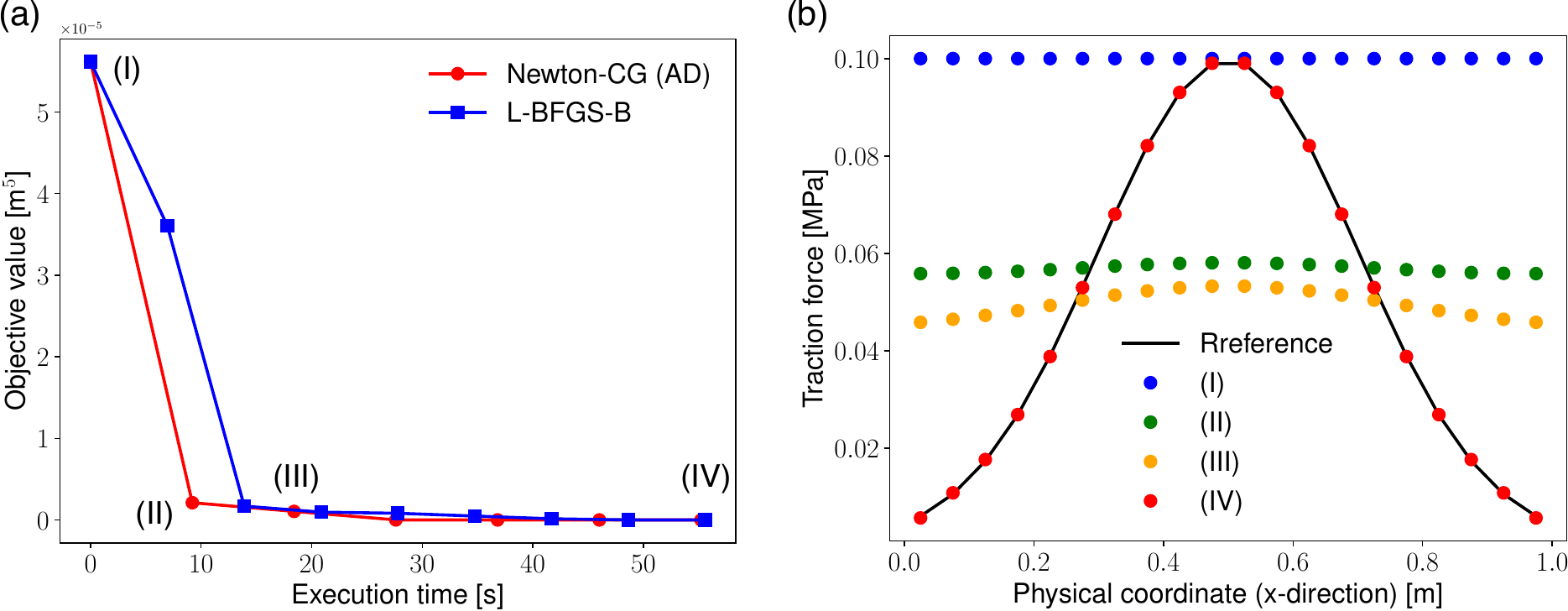}
\caption{Optimization results for the traction field identification problem. In (a), the objective value converges to close zero for both Newton-CG (AD) and L-BFGS-B. In (b), the traction fields at step (I), (II), (III) and (IV) are shown to compare with the reference traction field.}
\label{Fig:hyperelasticity_opt}
\end{figure}

The three optimizers L-BFGS-B, Newton-CG (FD) and Newton-CG (AD) are used to solve the PDF-constrained optimization problem, and the results are shown in Fig.~\ref{Fig:hyperelasticity_opt}-(a), where the objective value is plotted against the actual execution time, and each data point is recorded at every optimization iteration for both L-BFGS and Newton-CG (AD). 
Similar to the inverse optimization of the previous linear Poisson's problem, the result of Newton-CG (FD) is missing, since the optimizer does not converge and cannot give a valid result.
In contrast, Newton-CG (AD) converges and the optimization is successful.
This contrast further emphasizes the importance of obtaining accurate Hessian-vector products instead of approximate ones.
In Fig.~\ref{Fig:hyperelasticity_opt}-(a), we mark four representative iteration steps of Newton-CG (AD) from step (I) to step (IV), and show the corresponding traction force fields in Fig.~\ref{Fig:hyperelasticity_opt}-(b).
We see that the final traction field at step (IV) is close to the reference traction field, showing that the traction field identification task is successful.

\begin{figure}[H] 
\centering
\includegraphics[scale=0.65]{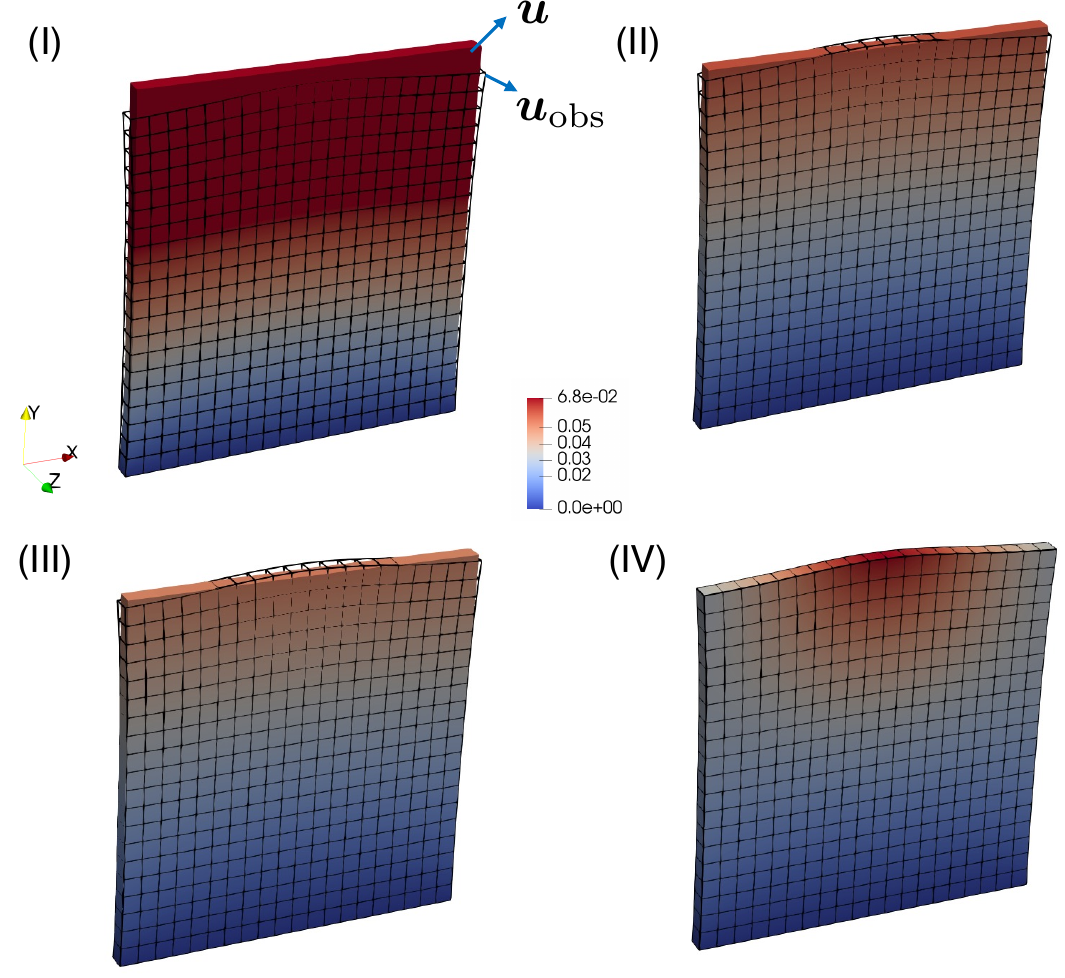}
\caption{Matching between the predicted displacement fields (colored plot) and the observed displacement field $\ub_\textrm{obs}$ (wireframe plot) at optimization steps (I) to (IV).}
\label{Fig:hyperelasticity_inv}
\end{figure}

To further show that the predicted displacement field indeed matches the observed displacement field $\ub_{\textrm{obs}}$, we show the predicted displacement fields at step (I) to (IV) in Fig.~\ref{Fig:hyperelasticity_inv}, along with the observed displacement field $\ub_\textrm{obs}$.
As shown, the predicted displacement field at the final step (IV) agree well with the observed displacement field $\ub_\textrm{obs}$.

In this numerical example, we conclude that both L-BFGS-B and Newton-CG (AD) work reasonably well, and Newton-CG (AD) is recommended because of its faster convergence to a low objective value.
Newton-CG (FD) fails to generate any useful results.

\subsection{Thermal-mechanical control}

\begin{figure}[H] 
\centering
\includegraphics[scale=0.5]{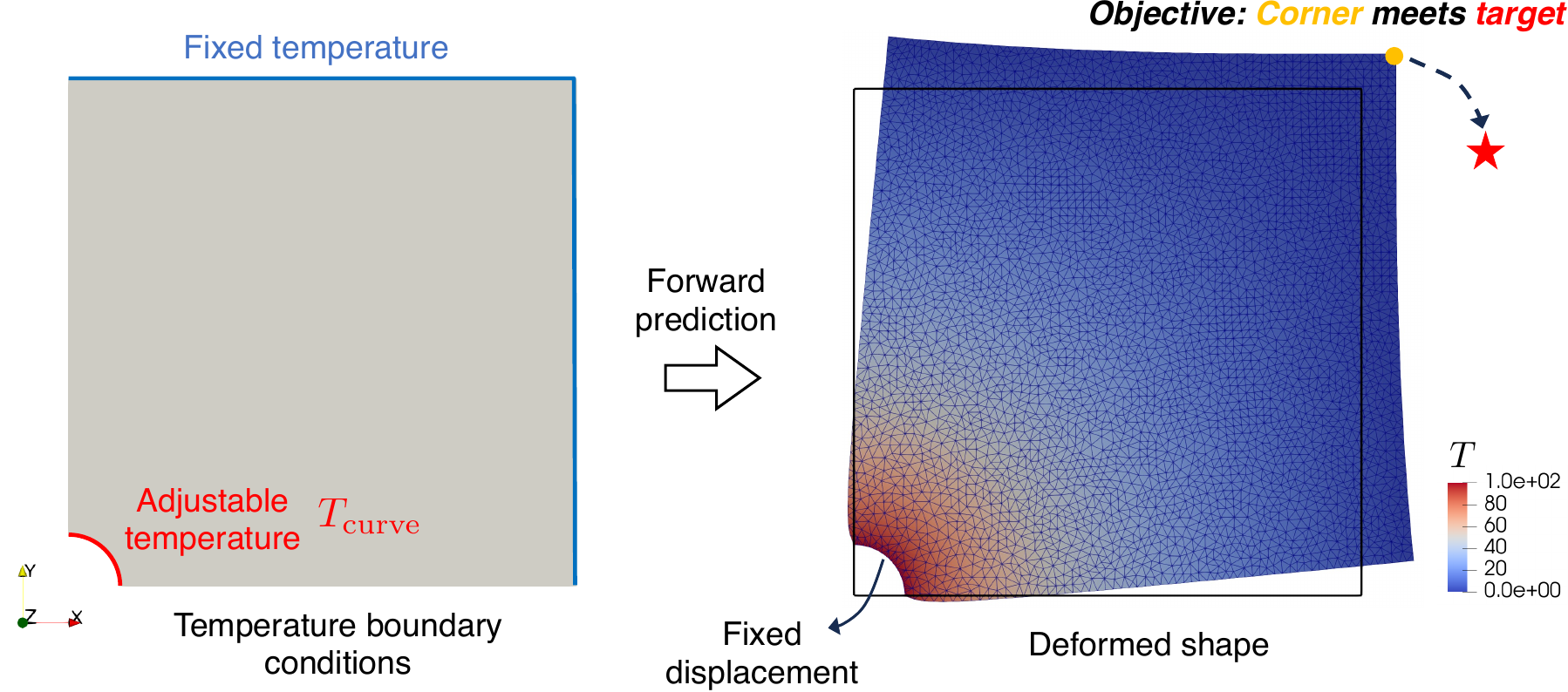}
\caption{The forward prediction of the thermal-mechanical problem. The deformation is scaled by a factor of 100. }
\label{Fig:thermal_mechanical_fwd}
\end{figure}

We consider a thermal-mechanical control problem, where the temperature defined on a curved boundary can be adjusted so that the thermal deformation of the object meets a desired goal.
As shown in Fig.~\ref{Fig:thermal_mechanical_fwd}, we set the objective to be that the top right corner of the 2D square-shaped object (yellow circle) reaches a prescribed target point (red star), realized by optimizing the temperature field $T_{\textrm{curve}}$ on the curved boundary.

Before we introduce the inverse control problem, we first describe the forward predictive problem.
Following the thermal-mechanical coupling problem introduced in the \texttt{FEniCS} tutorial~\cite{mitusch2019dolfin}, we consider a quarter of a square plate perforated by a circular hole, subject to a prescribed temperature field $T_{\textrm{curve}}$ at the curved boundary.
The governing equation for the steady state heat problem needs to solve for the temperature variable $T$
\begin{align} \label{Eq:strong_thermal}
   -\nabla \cdot (k \nabla T) &=  0  \quad \quad \quad \textrm{in}  \, \, \Omega, \nonumber \\
    T &= T_{\textrm{curve}}   \quad\textrm{on} \, \, \Gamma_{D_1, T},  \nonumber \\
    T &= 0  \quad \,\,\,\,\,\, \quad\textrm{on} \, \, \Gamma_{D_2, T},  \nonumber \\
     k \nabla T \cdot \nb &= 0   \quad \quad \,\,\,\,\, \,\textrm{on} \, \, \Gamma_{N, T},
\end{align}
where $k$ is the thermal conductivity, $\Gamma_{D_1, T}$ is the first part of the Dirichlet boundaries for temperature (marked as red in Fig.~\ref{Fig:thermal_mechanical_fwd}), $\Gamma_{D_2, T}$ is the second part of the Dirichlet boundaries for temperature (marked as blue in Fig.~\ref{Fig:thermal_mechanical_fwd}), and $\Gamma_N$ is the Neumann boundary.
The governing equation for mechanical equilibrium needs to solve for the displacement variable $\ub$
\begin{align} \label{Eq:strong_mechanical}
    -\nabla \cdot \cs = \boldsymbol{0} & \quad \textrm{in}  \, \, \Omega, \nonumber \\
    \ub = \boldsymbol{0} &  \quad\textrm{on} \, \, \Gamma_{D,\ub},  \nonumber \\
    \cs \cdot \nb = \boldsymbol{0}  & \quad \textrm{on} \, \, \Gamma_{N,\ub},
\end{align}
where $\cs$ is the Cauchy stress tensor, $\Gamma_{D,\ub}$ is the Dirichlet boundary (the curved part as shown in Fig.~\ref{Fig:thermal_mechanical_fwd}) and $\Gamma_{N,\ub}$ is the Neumann boundary.
The constitutive relation considering thermal effect is
\begin{align}
    \nonumber
    \boldsymbol{\sigma} & = \lambda \,\textrm{tr}(\boldsymbol{\varepsilon}) \, \boldsymbol{I}+2\mu\, \boldsymbol{\varepsilon} - \kappa\,T\,\boldsymbol{I} \\
    \boldsymbol{\varepsilon}&=\frac{1}{2}\Big(\nabla \boldsymbol{u}+(\nabla\boldsymbol{u})^T\Big),
\end{align}
where $\boldsymbol{\varepsilon}$ is the strain tensor, $\lambda$ and $\mu$ are the Lamé parameters, and $\kappa$ relates to the thermal expansion coefficient.
Note that the temperature variable $T$ considered in this numerical example is a relative temperature that refers to the temperature change from the ambient temperature.
The thermal and mechanical properties are isotropic and correspond to those of aluminum.

The weak form of this thermal-mechanical problem states that find $T$ and $\ub$ such that for any test function $\delta T$ and $\delta \ub$:
\begin{align} 
    r(T, \ub;  \delta T, \delta \ub) = \int_{\Omega} k \nabla T \cdot \nabla \delta T \, \textrm{d}\Omega + \int_{\Omega}  \cs : \nabla \delta\ub  \, \textrm{d}\Omega   = 0.
\end{align}

This thermal-mechanical problem uses a one-way coupling scheme in that the displacement $\ub$ depends on the temperature $T$, but not the other way around.
The forward prediction problem states that given the adjustable temperature boundary conditions $T_{\textrm{curve}}$, find the deformation of the object due to the thermal effect.

The problem is discretized using 7821 linear triangular elements in 2D for both $T$ and $\ub$.
The parameter field is the boundary temperature field $T_{\textrm{curve}}(\x)$.
The parameter vector $\thetab =[T_1,T_2,\dots,T_M]$ is the discretized version of the parameter field $T_{\textrm{curve}}(\x)$, where $M=26$ due to the 26 nodes on the curved boundary.
Since we are interested in the deformation of the object, the output solution vector $\y=[u_1, u_2,\dots,u_N]$ is the discretized version of $\ub(\x)$, where $N=4052\times 2$ due to the 4052 finite element nodes and the 2 degrees of freedom of displacement associated with each node.

The inverse control problem is defined as a PDE-constrained optimization problem:
\begin{align} \label{Eq:PDE-CO_thermal_mechanical}
    \nonumber &\min_{T, \ub, T_\textrm{curve}}  \Vert\ub_{\textrm{corner}} - \ub_{\textrm{target}}\Vert^2  
        + \frac{\alpha}{2} \int_{\Gamma_{N,T}} \Vert\nabla T_{\textrm{curve}}\Vert^2 \textrm{ d} \Gamma \\
    & \qquad \qquad \textrm{s.t.} \quad r(T, \ub;  \delta T, \delta \ub)=0, 
\end{align}
where $\ub_{\textrm{corner}}$ is the displacement of the top right corner that can be obtained from $\ub$ directly, $\ub_{\textrm{target}}=[0.001, -0.001]$ is the prescribed target displacement, and the regularization term encourages the smoothness the parameter field $T_{\textrm{curve}}$. 

\begin{figure}[H] 
\centering
\includegraphics[scale=0.5]{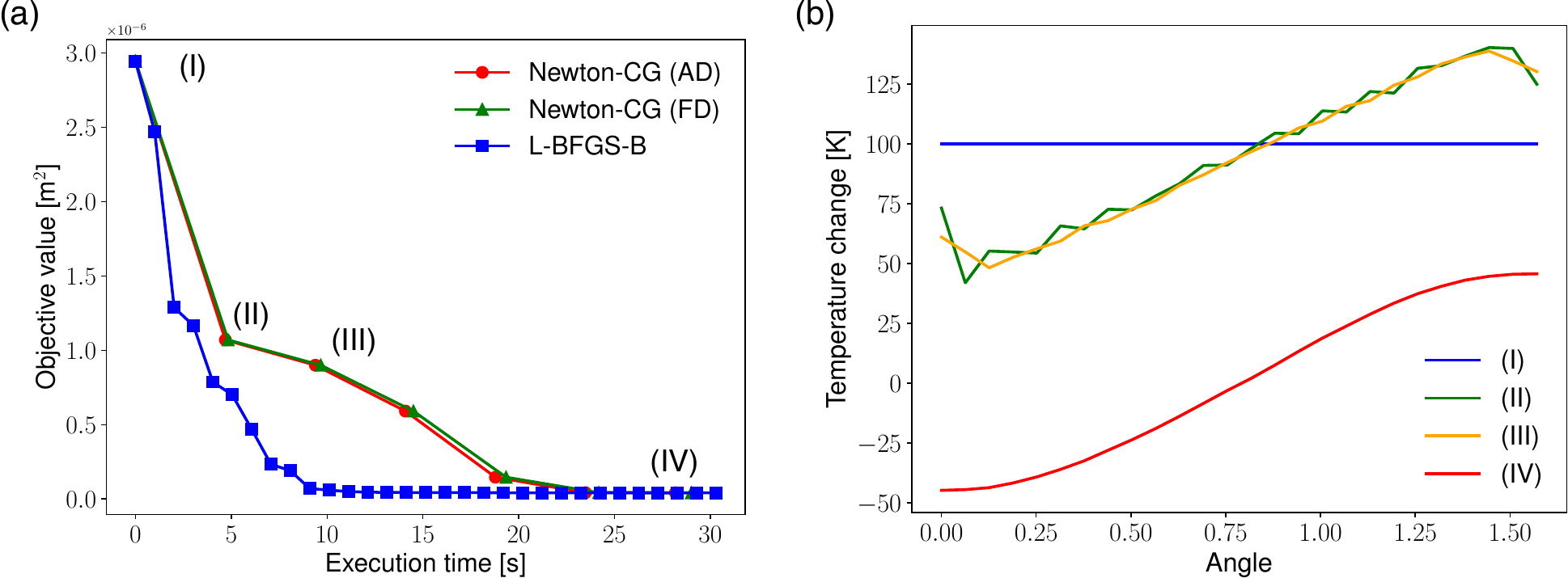}
\caption{Optimization results for the thermal-mechanical control problem. In (a), the objective value converges to close zero for Newton-CG (AD), Newton-CG (FD) and L-BFGS-B. In (b), the temperature parameter field $T_{\textrm{curve}}$ at step (I), (II), (III) and (IV) are plotted with respect to the polar angle.}
\label{Fig:thermal_mechanical_opt}
\end{figure}

The three optimizers L-BFGS-B, Newton-CG (FD) and Newton-CG (AD) are used to solve the PDF-constrained optimization problem, and the results are shown in Fig.~\ref{Fig:thermal_mechanical_opt}-(a), where the objective value is plotted against the actual execution time.
We mark four representative iteration steps of Newton-CG (AD) from step (I) to step (IV), and show the corresponding parameter field $T_{\textrm{curve}}$ in Fig.~\ref{Fig:thermal_mechanical_opt}-(b).
The red curve shows the optimized $T_{\textrm{curve}}$ which fulfills the optimization goal.

\begin{figure}[H] 
\centering
\includegraphics[scale=0.66]{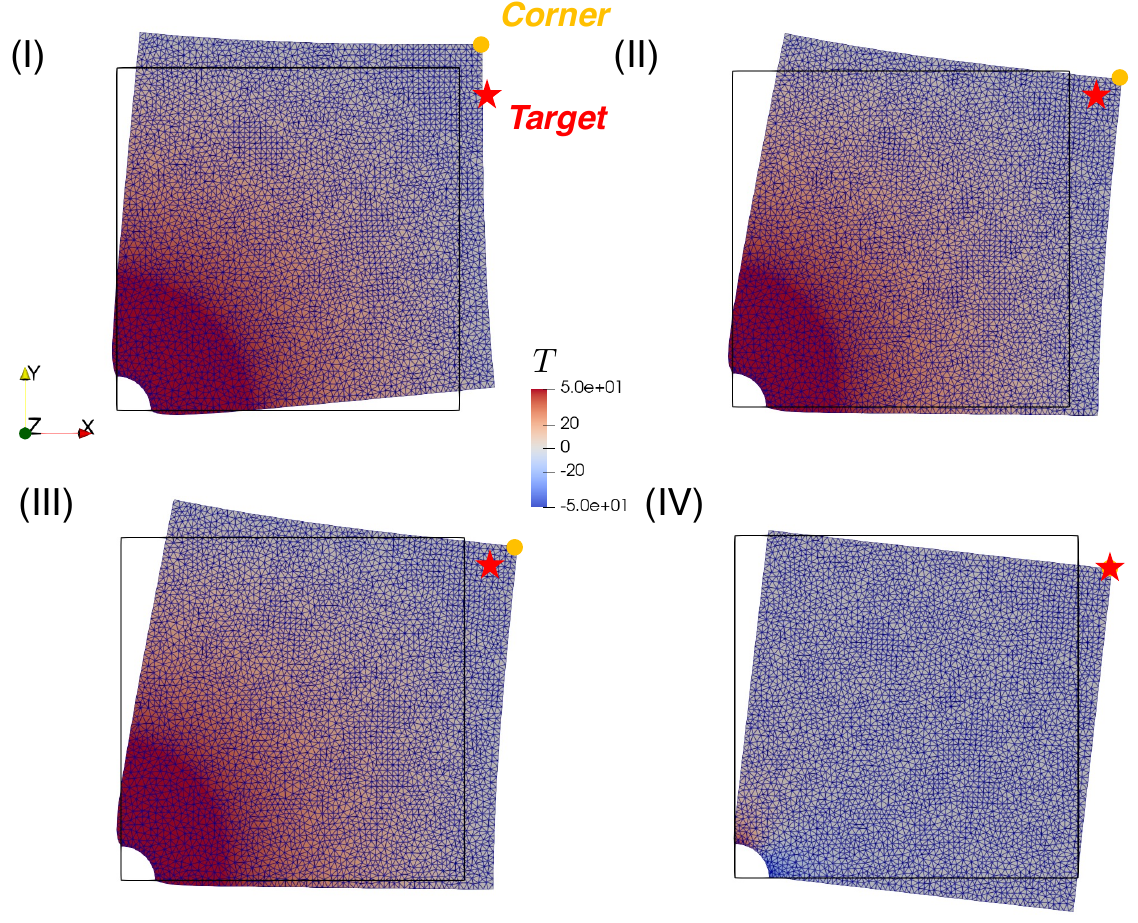}
\caption{The deformed object at optimization steps (I) to (IV), with corner displacement gradually matching target displacement. The deformation is scaled by a factor of 100.}
\label{Fig:thermal_mechanical_inv}
\end{figure}

To further show that the optimized parameter field $T_{\textrm{curve}}$ indeed leads to a deformation that achieves the target, we show the deformed object at step (I) to (IV) in Fig.~\ref{Fig:thermal_mechanical_inv}.
As shown, the corner displacement at the final step (IV) agrees well with the target displacement.

In this numerical example, we conclude that all the three methods Newton-CG (AD), Newton-CG (FD), and L-BFGS-B work reasonably well, and L-BFGS-B is recommended due to its faster convergence to a low objective value.
Newton-CG (AD) is slightly faster than Newton-CG (FD).

\subsection{Shape optimization}

\begin{figure}[H] 
\centering
\includegraphics[scale=0.55]{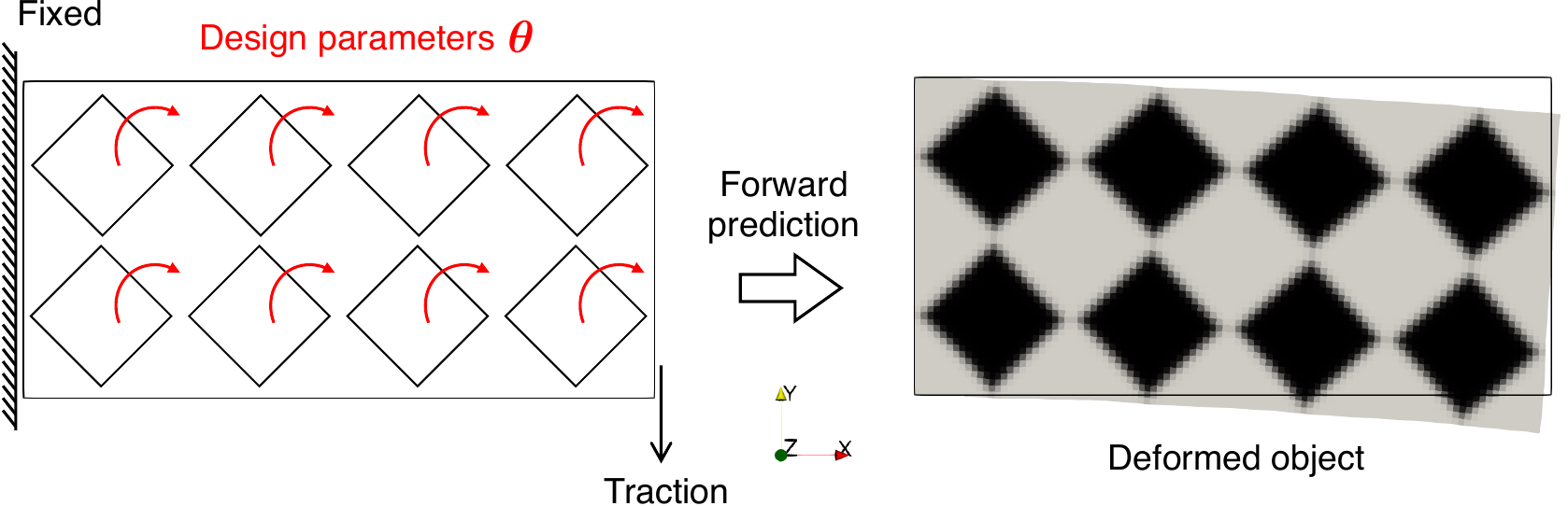}
\caption{The forward prediction of the shape optimization problem. }
\label{Fig:example_shape_fwd}
\end{figure}

Structural topology optimization~\cite{bendsoe2013topology} can be mathematically formulated as PDE-constrained optimization problems.
In this numerical example, we consider a related problem of structural shape optimization. 
As shown in Fig.~\ref{Fig:example_shape_fwd}, a 2D cantilever beam with eight square holes is fixed at the left side and subject to a shear force near the bottom of the right side.
Assume the manufacturing of the beam permits the rotation of these eight holes, and the goal is to minimize the compliance of the beam subject to the applied shear force.
The material model is the same as the previous example in Section \ref{Sec:traction}.
Therefore, we skip the strong form (similar to Eq.~(\ref{Eq:strong_hyperelasticity})) and the weak form (similar to Eq.~(\ref{Eq:weak_hyperelasticity})).

The problem is discretized using 5000 linear quadrilateral elements in 2D for the displacement solution $\ub$.
The design parameters $\thetab=[\theta_1,\theta_2,\dots,\theta_M]\in\mathbb{R}^M$ with $M=8$ implying the 8 rotational angles of the holes.
The output solution vector $\y=[u_1, u_2,\dots,u_N]$ is the discretized version of $\ub(\x)$, where $N= 5151\times 2$ due to the 5151 finite element nodes and the 2 degrees of freedom of displacement associated with each node.
The mapping from $\thetab$ to the density field $\rho(\x)$ commonly used in topology optimization relies on a sigmoid function to ensure smoothness for differentiability.

The inverse shape optimization problem is defined as a PDE-constrained optimization problem:
\begin{align} \label{Eq:PDE-CO_shape}
    \nonumber \min_{\ub,\thetab}   \int_{\Gamma_N} \tb\cdot\ub \textrm{ d} \Gamma \\
     \qquad \qquad \textrm{s.t.} \quad r(\ub;\vb)=0, 
\end{align}
where $\tb$ is the applied shear traction, and $r(\ub;\vb)=0$ is the PDE constraint in the weak form.

\begin{figure}[H] 
\centering
\includegraphics[scale=0.5]{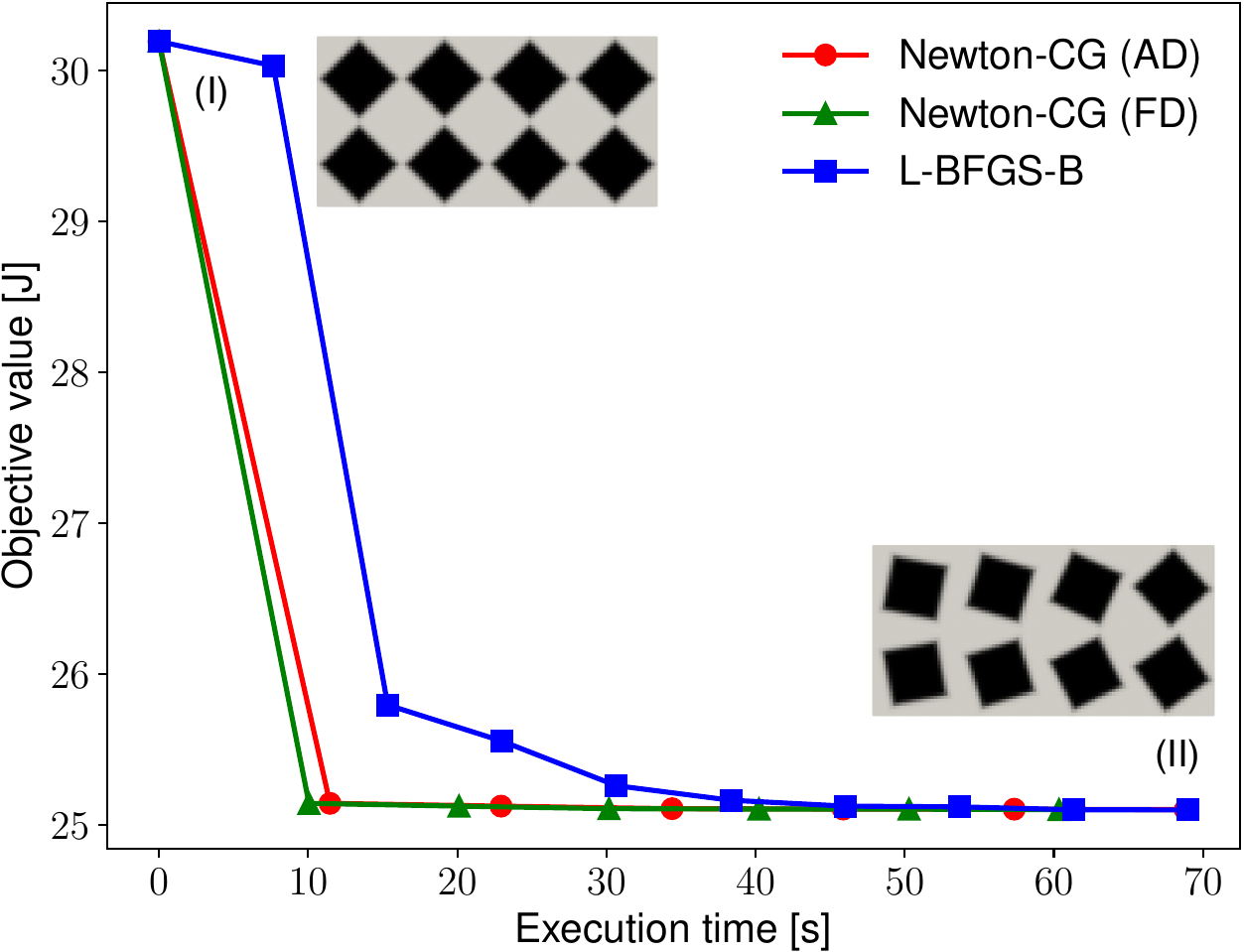}
\caption{Optimization results for the shape optimization problem. The configurations of the square holes at the initial step and the final step of Newton-CG (AD) are marked as (I) and (II). }
\label{Fig:shape_opt}
\end{figure}

The three optimizers L-BFGS-B, Newton-CG (FD) and Newton-CG (AD) are used to solve the PDF-constrained optimization problem, and the results are shown in Fig.~\ref{Fig:shape_opt}, where the compliance value is plotted against the actual execution time.
We mark the initial step and the final step of Newton-CG (AD) as step (I) and step (II), where the corresponding rotated square holes are shown.

In this numerical example, we conclude that all the three methods Newton-CG (AD), Newton-CG (FD), and L-BFGS-B work reasonably well, and Newton-CG (FD) is recommended due to its faster convergence to a low objective value.

\subsection{Discussions}

Based on the four numerical examples, we offer suggestions for choosing the suitable optimizers. 
In general, Newton-CG (AD) and L-BFGS-B both work well for all the examples, while Newton-CG (FD) does not converge for two out of four examples. 
For robustness, Newton-CG (AD) and L-BFGS-B are recommended. 
In terms of efficiency, Newton-CG methods converge faster for the two nonlinear examples (``traction force identification'' and ``shape optimization'') because of the precise information of second-order derivatives in the form of Hessian-vector products, overcoming complex curvature efficiently.
In contrast, L-BFGS-B converges faster for the two linear examples (``source field identification'' and ``thermal-mechanical control'') since it exploits simplicity and fast approximations for linear objectives.
In the cases where Newton-CG (FD) can converge, the execution time is similarly to Newton-CG (AD), which agrees with our previous analysis.
Newton-CG (FD) has the advantages of being simple, bypassing the need to implement the exact Hessian-vector products, while Newton-CG (AD) proposed in this work offers a robust scheme, particularly useful for the cases where Newton-CG (FD) fails to converge.

\section{Conclusions}
\label{Sec:conclusions}

Our work advances the integration of second-order implicit differentiation into finite-element-based differentiable physics, addressing a pressing need in PDE-constrained optimization. 
We present a framework for computing Hessians under the discretize-then-optimize paradigm, generalizing adjoint methods to second-order derivatives. 
By leveraging \texttt{JAX}’s automatic differentiation primitives (JVP/VJP), we implement Hessian-vector products, validating their accuracy against finite difference benchmarks. 
Four case studies, ranging from linear elasticity to nonlinear hyperelasticity, demonstrate the practical value of second-order derivatives. 
For nonlinear inverse problems, such as nonlinear traction identification and shape optimization, Newton-CG with exact Hessians achieves superior convergence compared to first-order methods, while L-BFGS-B remains effective for linear tasks. Crucially, our approach avoids the instability and inaccuracy of finite difference Hessian approximations, enabling robust optimization for complex physical systems. 
Future work could extend this framework to exploit sparsity in large-scale problems.
By providing open-source tools for second-order differentiable physics, this research empowers computational scientists to tackle high-dimensional design and control challenges with better robustness and efficiency.

\section{Acknowledgements}

The author gratefully acknowledges the Research Grant Council of Hong Kong for support through the ECS project (Ref No. 26205024).

\bibliographystyle{unsrt}
\bibliography{refs}


\end{document}